# Electricity storage requirements to support the transition towards high renewable penetration levels – Application to the Greek power system

Georgios N. Psarros* & Stavros A. Papathanassiou

*School of Electrical & Computer Engineering, National Technical University of Athens, Greece*

## Abstract

This paper investigates the electricity storage requirements to support the transition towards a high renewable energy source (RES) penetration in a cost-optimal manner. The achieved reduction of renewable energy curtailments and the decrease in the total generation cost of the system are quantified against a counterfactual scenario without storage. A methodology is presented to determine the optimum mix of short- and medium-duration storage needed to support system operation at increased RES penetration levels, using the mixed integer linear programming mathematical optimization. The Greek power system serves as a realistic study case, in its planned development for the year 2030, with a targeted annual RES energy penetration in the order of 60%. Li-ion batteries and pumped-hydro are selected as the representative technologies to include in the storage mix, assuming energy-to-power ratios of up to 6 hours for the former and 10 hours for the latter. It is shown that the introduction of a suitable mixture of storage facilities may improve renewable energy integration and, at the same time, reduce system cost to the extent that entirely compensates for the full cost of storage, thus allowing for a net economic benefit for the system. The optimum storage portfolio for the study case system and the targeted RES penetration level combines 2-h batteries and 6-h pumped-hydro stations, with an aggregate capacity of new facilities between 1250 MW and 1750 MW, on top of the existing 700 MW of open-loop pumped hydro plants. The optimum storage requirements vary with the targeted RES penetration and with the balance of RES technologies in the generation mix, particularly the level of PV integration.

**Keywords:** *cost-benefit analysis; electricity storage; storage requirements; value of storage*

## 1. Introduction

The decarbonization of the electricity sector involves the transformation of the entire generation system, with increasing reliance on renewable energy sources (RES) [1]. The intermittency of traditional stochastic RES and the inevitable errors in their prognosis create increased flexibility requirements, impacting the economic operation and security of the system [2–4], while curtailments in RES production constrain the exploitation of available potential under high RES penetration rates, [5]. Within this context, energy storage is an enabling factor to achieve the fast and effective decarbonization of energy systems, relying primarily upon stochastic RES production, [6–10].

The value of energy storage has been extensively analyzed in the literature ([11–18]). Available studies deal with the operating principles, characteristics, performance indicators of various storage technologies [12–15], storage investment cost trends [12,14–17], as well as the integration of storage in the electricity markets of the future [11,16,18]. Pumped-hydro (PHS) and electrochemical storage (batteries) are the principal technologies today and in the near future ([19–21]). PHS is still the dominant storage technology worldwide,

* Corresponding Author
e-mail addresses: gpsarros@mail.ntua.gr (G. N. Psarros), st@power.ece.ntua.gr (S. A. Papathanassiou)

accounting for 130 GW of installed capacity ([13,14,16,22]), while Li-ion batteries are the emerging technology, experiencing a fast cost decline ([17,23]) due to their application in electromobility.

Storage stations with a large energy capacity, such as PHS, offer energy arbitrage services, contribution to system resource adequacy ([24,25]), and mitigation of renewables curtailment [26]. Fast-response storage stations, such as Li-ion batteries, may provide increased operating reserves and flexibility ([26]), substituting thermal production and thus creating additional room to accommodate renewable capacity ([27]). Other long-duration storage technologies, such as hydrogen, ammonia, ethanol, etc., possibly subject to seasonal management principles, can also play a significant role in net-zero power systems in the long run, especially as the interlink for sector coupling applications ([28–30]). However, such emerging storage solutions are yet to become economically efficient and technologically mature for wide-scale application in power systems ([31–33]).

Until recently, literature was mainly investigating PHS as the only viable storage technology alternative in the energy mix ([34–37]) to support the transition towards low carbon emissions systems. More recent studies are focusing on battery energy storage systems (BESS), in line with the escalating investment interest in this technology [38–42], or on combinations of storage technologies, usually including both PHS and BESS [26,43,44]. The value of storage is also investigated in technology-agnostic approaches [45–49], revealing the benefits arising from the generic functionality of storage, where technology-specific characteristics are ignored or exist only related to the investment cost, such as in [45]. On the other hand, technology-specific modeling reveals the pros and cons of the examined storage alternatives. A variety of BESS technologies are considered in [26], while PHS with or without pre-existing water reservoirs are accounted for in [44]. The value of PHS technology is highlighted in [37] and [26] in supporting increased RES penetration levels, above 60% and 80% respectively. In [26], BESS of short duration (up to 3-h) are also found vital for procuring the required reserves to the grid.

Studies addressing the appropriate energy storage mix typically employ capacity expansion modeling (CEM) or day-ahead scheduling (DAS) simulation methods. The CEM approach optimizes system operation over a long-term horizon, typically extending over one or more years, simultaneously determining new investments in generation, storage, and possibly transmission, taking into account the investment and operating cost of the system. In CEM, storage is sized either assuming a predefined duration (i.e., a fixed energy-to-power ratio, [38,43,49]) or independently optimizing the power and energy capacity of system storage, assuming distinct costs for its energy and power components ([26,42,44]). CEM approaches may determine the optimum storage mix, involving multiple storage technologies, while they can manage long-duration and seasonal storage, such as power-to-gas technologies ([26]). On the other hand, such models suffer from increased computational complexity due to their extended optimization horizon, often resorting to simplifications to reduce the number of variables and constraints of the problem ([42]). For instance, clustering techniques for power plants and preselection of typical days ([7,44]) may be employed, and system management and operational constraints may be over-simplified in the attempt to maintain a balance between detail and feasibility of the solution in acceptable timescales.

The DAS-based models follow a fundamentally different approach, optimizing system operation over a limited look-ahead horizon that usually extends from 24 to 36 hours ([34,37,39,46]). Their main target is to capture in detail the impact of storage on system operation, adopting a finer and ultimately more realistic representation of actual unit commitment and economic dispatch (UC-ED) practices, especially as they relate to market conditions. To determine the value of storage, several studies rely only upon the results of a single execution of DAS for an indicative day ([34]) or a cluster of days ([47]), then extrapolating to longer time scales, inevitably missing essential aspects related to the seasonality of load demand, renewable resource availability and their correlation. A more sophisticated approach would involve the consecutive execution of DAS over an entire year, reproducing the operation of system assets without such compromises. To quantify the value of storage, simulations of system operation with and without storage are typically employed. DAS-based models are valuable tools to evaluate the impact of storage, especially when dealing with short-duration assets, addressing flexibility needs of the system.

This paper addresses the fundamental question of quantifying storage requirements of power systems transitioning towards increased renewable energy shares. Optimal storage needs are quantified in terms of power and energy capacity, with both PHS and BESS technologies being considered for individual or combined application. The analysis employs a DAS-based algorithm with hourly granularity, consecutively executed to simulate the annual operation of the system. The model adopts a cost-optimal objective and includes conventional thermal, hydro, RES generation, system-level storage, and demand-side response services of limited flexibility. A DAS-based analysis methodology has been selected over the CEM alternative for the following reasons:

- Firstly, it is suitable for application as the only variables to be defined by the analysis are the power and energy capacity of storage assets, with the installed capacities of thermal units and intermittent RES being already defined.
- It affords a more detailed representation of system operation, especially regarding flexibility requirements and their correlation with BESS resources delivering an array of services beyond energy shift ([44]).
- It employs a high-fidelity model to simulate the generation system without resorting to simplifications, such as clustering techniques, which may compromise the validity of the results ([50,51]). The inclusion of constraints essential to capture the management of flexibility assets, such as reserves requirements per type and direction, unit start-stop functionality, etc., are easily incorporated in a DAS model. On the other hand, that would not be feasible in the context of a CEM optimization model.
- It scans the entire space of potential storage configurations, providing a comprehensive overview of the benefits derived from the introduction of different storage mixtures, detecting also feasible combinations leading to similar system benefits. With a traditional CEM algorithm, a single solution would have been obtained.

Optimization is formulated as a mixed integer linear programming (MILP) problem, tracing a cost-optimal solution from a system-level perspective, without introducing market-level details and related distortions. The optimal storage portfolio is selected on the basis of the achieved improvement in the annual generation cost of the system, with respect to the case without storage, including the operating cost of generation assets, as well as the annualized fixed capital cost of storage investments to ensure their feasibility. In this evaluation, the contribution of storage to resource adequacy is also considered ([25,52]), properly monetized.

The power system of Greece is used as a case study, adopting a RES penetration target of around 60%, as foreseen in the National Plan for Energy and Climate (NPEC) for 2030, [43]. The generation portfolio of the Greek system in the mid-term horizon to 2030 is well-defined in the NECP, with storage being the main asset yet to be identified. The analysis includes the principal storage technologies available for large-scale deployment in the near term, i.e., BESS and PHS, while storage needs are also evaluated at different RES penetration levels and mixtures.

The remaining of this paper is organized as follows: Section 2 presents the evaluation methodology adopted. Section 3 describes the study case system and the assumptions adopted for the simulation. Section 4 presents the main analysis results regarding the optimum storage portfolio for a 60% RES share, while Section 5 includes a sensitivity analysis for lower RES penetration levels and a diversified RES mix. Conclusions are summarized in Section 6. Appendix A includes an analysis regarding the contribution of storage to resource adequacy. System characteristics and investment cost data are presented in Appendix B. Appendix C incorporates a description of the mathematical formulation of the UC-ED problem.

# 2. Methodology

## 2.1. Economic evaluation of storage development scenarios

The system-level economic benefits derived from the introduction of a particular storage configuration are evaluated against a counterfactual scenario without storage. For the definition of the counterfactual scenario, two approaches can be considered, similar in principle to [53]:

- The *business as usual* (*bau*) alternative: In this case, the counterfactual scenario consists of the same system configuration, without the considered storage assets; the latter are simply omitted, without being substituted by other assets or means to the replenish their contribution to capacity adequacy. This is a valid hypothesis for systems not experiencing resource adequacy issues, in which case the benefits resulting from the introduction of storage do not include the monetized value of their capacity contribution.

- The *do-minimum* alternative: Here, the contribution of storage to resource adequacy is essential for the system, hence storage assets are substituted in the counterfactual scenario by additional dispatchable generation providing a commensurate capacity contribution. In this paper, open-cycle gas turbines (OCGT) are considered for this purpose, being the alternative with the lowest investment cost, greatest ease of deployment and minimum impact on system operation.

Fig. 1 illustrates the methodology applied to evaluate the system benefit derived for each storage development scenario, against the *bau* and *do-minimum* counterfactual. The annual economic benefit from the introduction of a particular storage configuration is first calculated as the difference in system operating cost[†] against the baseline scenario without storage (*bau* alternative). This difference represents savings in the variable cost of the system achieved through the considered storage mix, primarily via improved renewable energy exploitation and more efficient operation of conventional generation. These savings are reduced by the annualized fixed capital and operating cost of storage investments, calculated according to Appendix B, to determine the net benefits against the *bau* counterfactual scenario (green frame in Fig. 1). The total economic benefit achieved by a particular storage configuration, evaluated against the *do-minimum* counterfactual scenario (yellow frame in Fig. 1), includes the value of storage contribution to resource adequacy, which is quantified as presented in Appendix A for energy-limited storage assets. This value is considered to be equal to the avoided annual cost of a conventional generation solution, in our case OCGTs, providing the same capacity value.

[†] The system operating cost of each scenario consists of the variable operating costs of thermal units, including (a) fuel costs, (b) start-up/shut-down costs, and (c) $CO_2$ emissions rights, and the cost attributed to the net energy imports by neighboring countries.

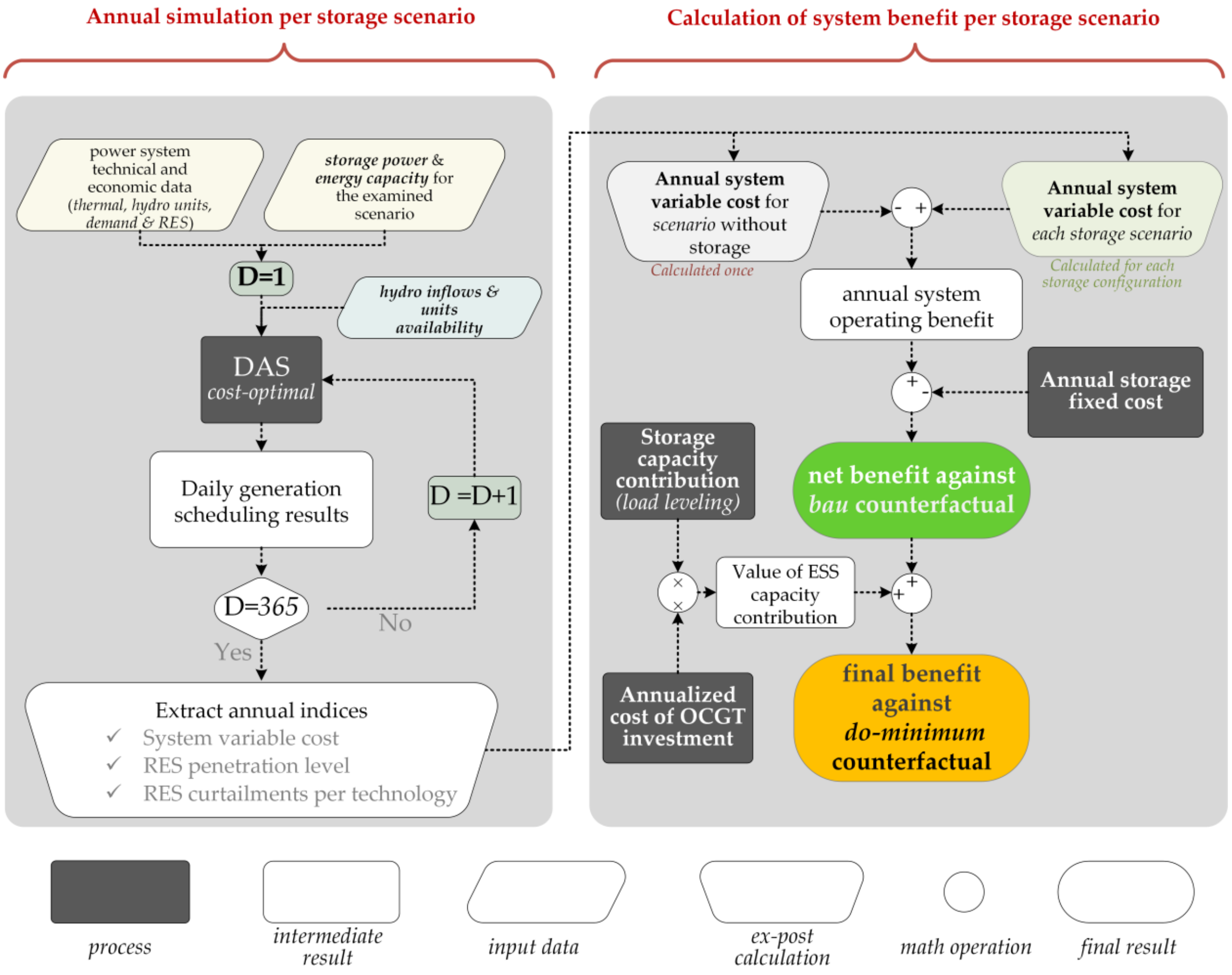

**Fig. 1:** Visualization of cost-benefit analysis conducted for each storage scenario. Benefits are assessed against a counterfactual scenario without storage.

At this point, it should be stressed that, besides the economic benefits quantified by the proposed methodology, storage will also provide additional services to the system, which are not evaluated and monetized in the paper; primarily technical services, including, inter alia, contribution to system inertia, voltage support and stability, black-start capability, network congestion management, fault level contribution ([54–58]), etc. If these services were properly quantified and monetized, the resulting overall value of storage would have been substantially higher than calculated in the paper.

## 2.2. Fundamentals of DAS-based simulation & mathematical formulation

To evaluate the annual variable cost of the system and thus the savings achieved through the introduction of storage, a DAS-based UC-ED model is employed, with a 24-h look-ahead horizon and an hourly time step, consecutively executed over the course of a year. Note that the 1-hour granularity used in this paper is typical for system planning and cost-benefit studies ([26,38,42–44,49]), even though the actual market resolution may be higher (e.g. 15 min for European balancing markets - [59]). The initial conditions for all generation and storage assets at the beginning of each day are given by their operating status at the end of the previous day. This process is described in the left-hand side flowchart in Fig. 1.

The objective of the optimization performed within the UC-ED model is to minimize the operating cost of conventional generation and maximize RES absorption, while concurrently fulfilling reserves requirements of the system. Such objectives and constraints may conflict, especially in high RES penetration conditions. Thus, the optimal solution will typically involve a certain level of RES energy curtailments and possibly the compromised fulfillment of operating reserves requirements.

The adopted system simulation methodology is a purely cost-optimal one, valid under the assumption of a well-functioning and fully competitive market, where all participants are bidding based on their marginal cost to enjoy dispatching priority. Market particularities, distortions due to strategic behaviors of participants with market power and other similar effects are not modeled in this study. Additionally, perfect foresight is assumed for renewable production in DAS. Storage primarily serves as a flexibility asset, being dispatched to optimize system benefits without submitting energy production/absorption bids. Technical and economic data of conventional, hydro, RES, and storage units are taken into account, while cross-border interconnections with neighboring countries are modeled based on their net transfer capacity.

The optimization problem is built upon the state-of-the-art MILP technique, incorporating linearized approximations of all non-linear constraints. The mathematical representation of the MILP optimization problem is briefly described in Appendix C. The algorithm is implemented in GAMS using the CPLEX optimizer, while the initialization of the problem is performed in MATLAB. For the analysis of this paper, over 150 annual scenarios have been analyzed. The computation time for a 24-h execution of the UC-ED lies between 30 s and 5 min and involves 205,290 variables.

# 3. Study case system, assumptions & scenarios

Energy storage portfolio requirements are determined using the Greek power system as a case study, in its anticipated development in year 2030, as stipulated in the approved NECP ([60]). In the decade 2020-2030, the Greek generation system is going through a significant transformation, involving the complete phase out of the so far dominant lignite-based generation fleet, accompanied by a great expansion of RES generation technologies, mainly wind farms (WFs) and solar PVs, targeting a share of renewable energy in electricity in excess of 60% by year 2030.

Annual energy demand in 2030 is anticipated at ~61.8 TWh, including final demand and electrical system losses, with a peak load of 11 GW. Installed RES capacities will comprise 7.0 GW of WFs and 7.7 GW of PVs, with an annual capacity factor of 28.5% and 18.1% respectively. Out of this capacity, 3 GW of WFs and 4.2 GW of PVs are considered to be capable of curtailment. An additional 1.0 GW of dispatchable RES technologies is also foreseen, including biomass/biogas, run-of-the-river small hydro, geothermal and concentrating solar thermal stations. The generation fleet also comprises 7.5 GW of Combined Cycle Gas Turbines running on natural gas with a variable cost ranging from 65 to 90 €/MWh, and 3.5 GW of large hydroelectric plants with reservoirs and natural inflows. This capacity incorporates 0.7 GW of open-loop hydro pumped storage stations already in operation. Technical characteristics of conventional generators are in line with [61]. A detailed description of power system assumptions is presented in Appendix B.

The Greek system is interconnected with five neighboring countries, with a cumulative net transfer capacity of 2.1/1.5 GW (import/export, respectively). As far as system reserves requirements are concerned, three types of operating reserves are considered; the Frequency Containment Reserves (FCR) and the automatic & manual Frequency Restoration Reserves (aFRR & mFRR). FCR requirements equal to 60 MW, symmetrical in both directions. Maximum aFRR up/down requirements reach 1000/500 MW respectively, while mFRR up/down requirements are in the order of 2000/1000 MW.

To quantify the mid-term electricity storage needs of the Greek system, a wide range of BESS and PHS configurations are analyzed, introduced either as single technologies or combined in storage mixtures. The capacities of BESS and PHS vary between 0 and 2000 MW per technology, while the cumulative capacity of BESS and PHS stations varies in the range 1000 - 2000 MW with a 250 MW increment. BESS energy capacities evaluated are 2, 4 and 6 equivalent hours at rated power, while PHS stations of 6-h, 8-h and 10-h are considered

The roundtrip efficiency for PHS and BESS is set to 70% and 80% respectively‡. BESS contribute to the provision of all types of reserves, while PHS plants are excluded from the provision of FCR, as is the typical case with hydro stations in Greece. For storage to participate in reserves provision, sufficient energy reserves need to be maintained, here assumed equal to 15-min of operation at rated output power. For BESS stations, a permissible state of charge variation between 15% and 95% is assumed for normal operating conditions, without minimum loading restrictions (0 to ±100% of rated power capacity) and seamless transition between charging and discharging states, thus allowing enhanced reserves provision (in principle, up to twice their power capacity). PHS are assumed to present a minimum loading restriction of 25%/10% of their nominal power in pumping/generation mode.

# 4. Results

System operation without the introduction of new storage, besides the already existing open-loop PHS plants, leads to an annual RES penetration of 60.09% of demand, with a curtailment level of 4.2% and 2.6% for PVs and WFs subject to power limitations, respectively. This holds both for the *bau* and the *do-minimum* cases, as OCGTs considered in the latter only contribute to capacity adequacy without being dispatched because of their high variable operating cost (effectively being strategic reserve units).

## 4.1. Benefits through PHS storage

As shown in Fig. 2, the introduction of sufficient PHS capacity can significantly reduce renewable energy curtailments, eventually to levels below 1.2% and 0.8% for PVs and WFs, for PHS configurations over 1500 MW. It is also evident that PHS energy capacity is not as important in mitigating curtailments. For instance, assuming a PHS capacity of 750 MW and 6-10 h duration, leads to curtailments varying by only 0.1%-0.2%. This can be attributed to the following factors:

- RES over-generation predominantly occurs in midday hours of high PV production, especially in tandem with increased wind generation. Thus, curtailments take place within a narrow window lasting only a few hours around mid-day and are characterized by high peaks rather than prolonged durations and energy content. In such cases, storage durations beyond 6-h hardly contribute to a further reduction of curtailments, while storage power may be more significant for RES surplus exploitation. This is confirmed in Fig. 3, where PHS of different storage durations (6-h, 8-h and 10-h) but the same power capacity (750 MW) lead to almost identical curtailments. Conversely, as shown in Fig. 4, when PHS capacity is increased from 750 MW to 1500 MW, maintaining the same 8-h duration, RES curtailments are substantially reduced.

‡ A roundtrip efficiency of 70% is in line with data available for existing and planned PHS projects in Greece ([65]), even though it is relatively pessimistic given the range of efficiencies quoted in the literature for large PHS plants ([66]). Similarly, the 80% roundtrip efficiency assumed for BESS facilities lies in the lower end of the efficiency range of Li-ion battery systems ([63]), but it is supported by results obtained from real-world applications ([67]).

- When the system operates at high RES penetration, it will require increased amounts of active power reserves to deal with renewable intermittency. Hence, a significant amount of the available PHS power is allocated for reserves provision, leaving only a fraction of the power capacity available to perform energy arbitrage. Therefore, storages of a higher power capacity will possess the margin to perform arbitrage more systematically and therefore mitigate RES curtailments more effectively in terms of reserves provision, while operation leaves greater power margin for arbitrage after reserves allocation to the station. This can be observed in the first days in Fig. 4 (a)-(b), where the larger station is capable of arbitrage functionality, not afforded by the smaller plant.
- Energy arbitrage is generally performed when a significant price spread exists in the day ahead market, sufficient to cover the roundtrip losses of the storage systems; for PHS, whose losses are assumed equal to 30%, a selling price at least 43% higher than the buying price is required. Such price differences are limited to only a few hours per day, usually during RES overgeneration conditions leading to curtailments and near-zero prices. This is confirmed in Fig. 5, presenting the average daily energy transactions of PHS stations, expressed in hours of equivalent operation at rated power. It is observed that average durations range between 2-h and 4-h, therefore storages of a high energy capacity are under-utilized.

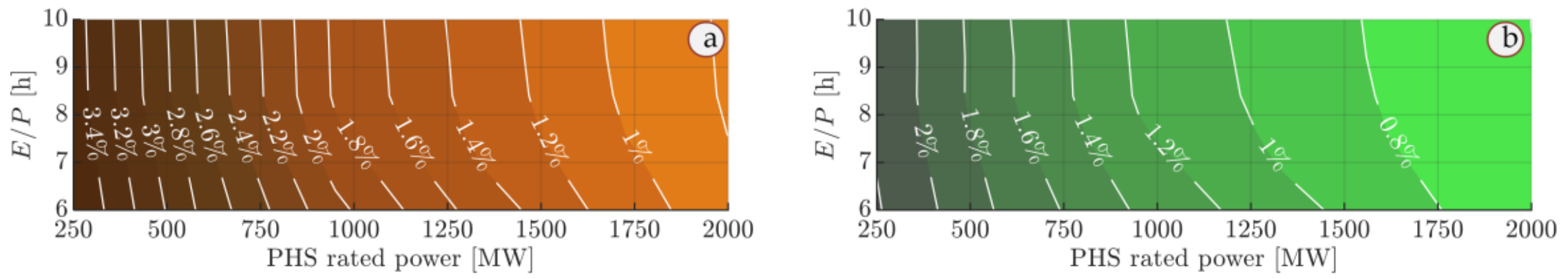

**Fig. 2:** Annual RES curtailments for various PHS configurations (power capacities and reservoir durations), in % of available RES energy, for (a) PV stations and (b) WFs.

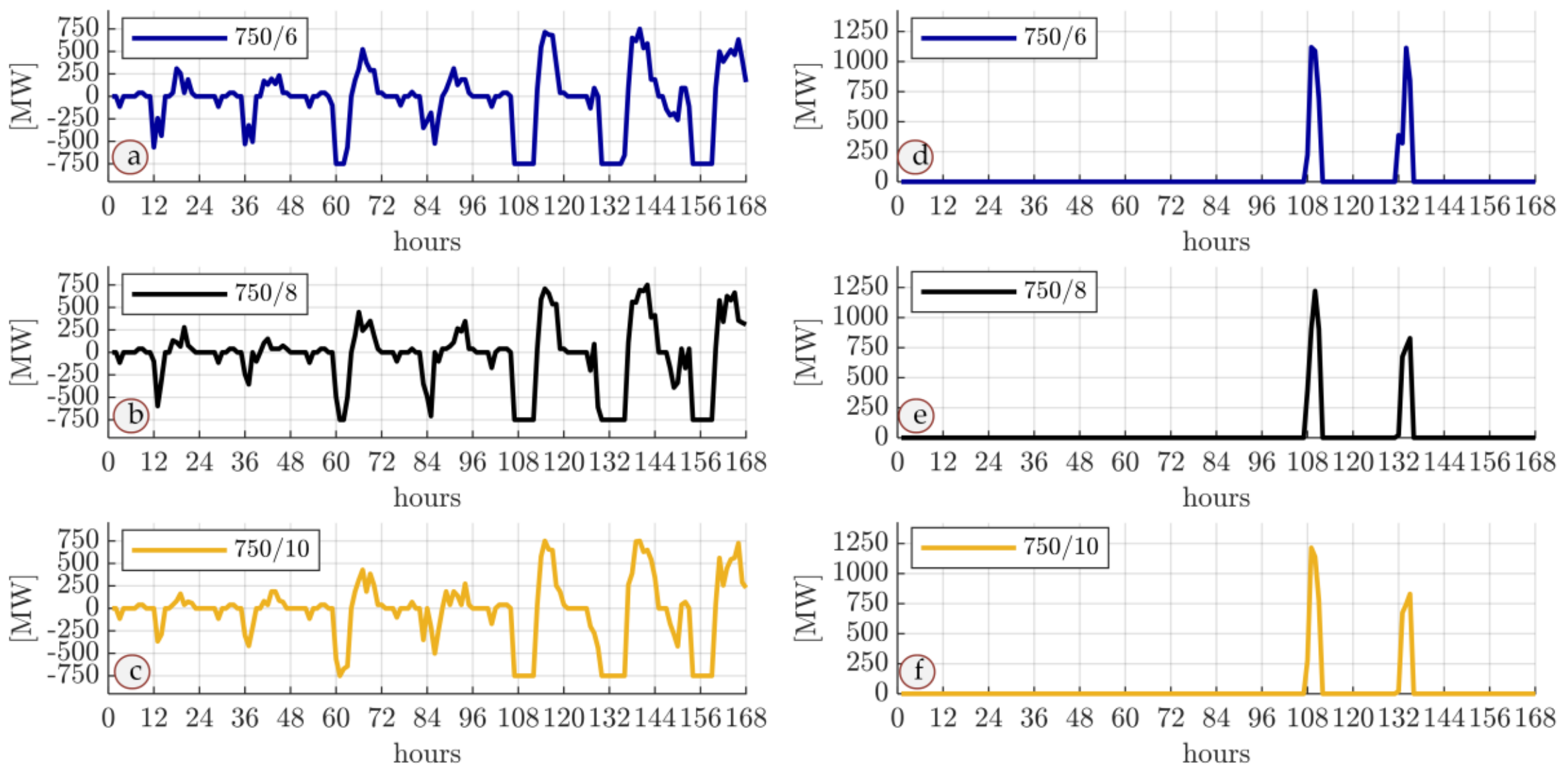

**Fig. 3:** (a)-(c) Operating profiles of three PHS configurations, having the same power capacity but different durations. (d)-(f) Respective renewable energy curtailments of the system.

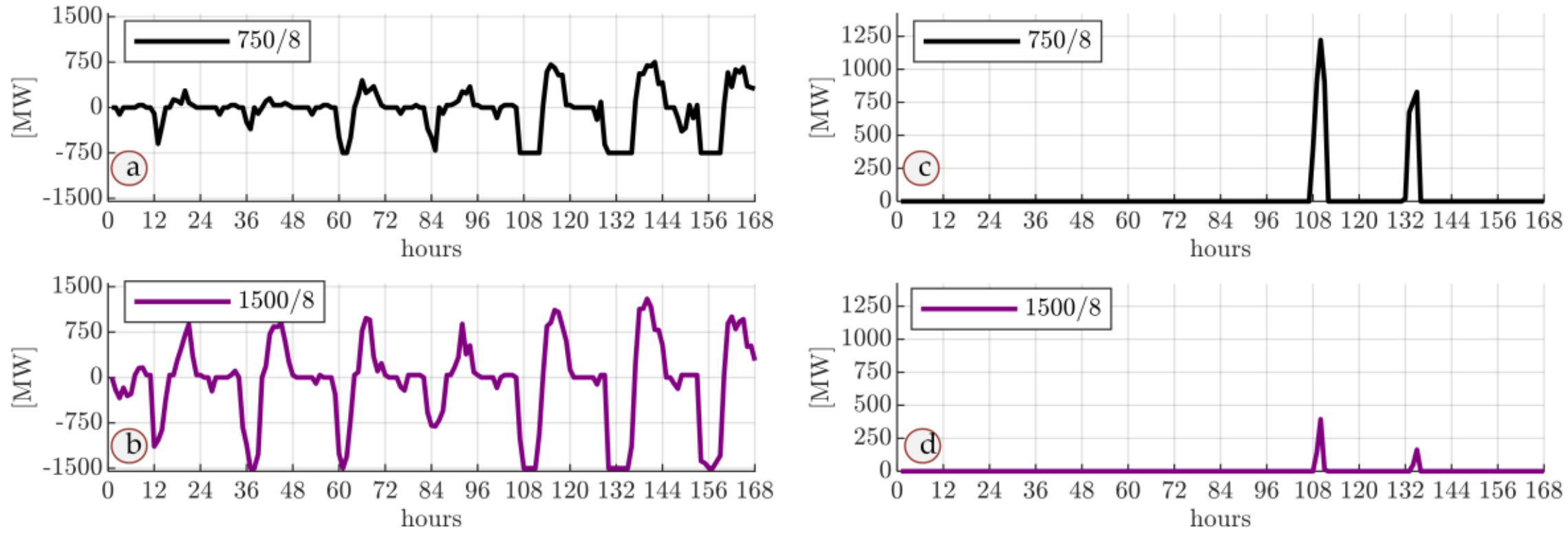

**Fig. 4:** (a)-(b) Operating profiles of two PHS configurations, having the same duration but different power capacities. (c)-(d) Respective renewable energy curtailments of the system.

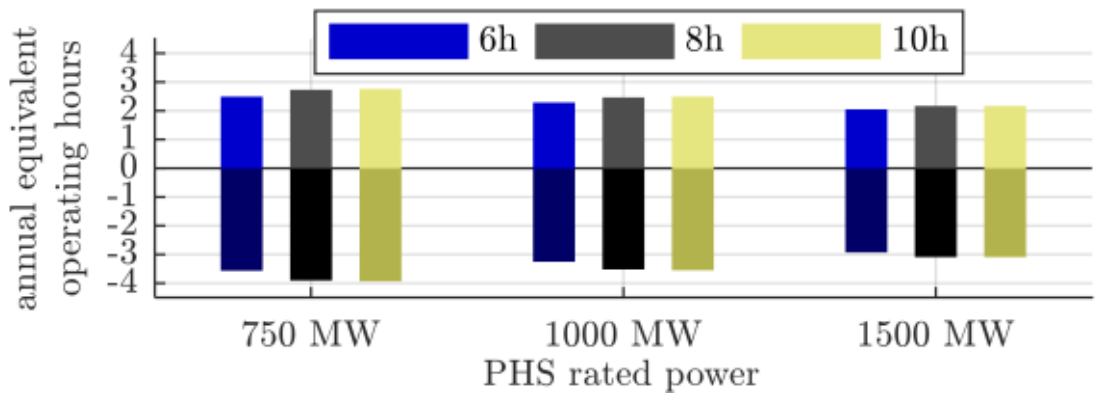

**Fig. 5:** Average daily energy transactions for different PHS configurations, expressed in equivalent hours at rated power. Negative values denote charging, positive values signify discharging.

Fig. 6 (a) presents the reduction in system cost achieved through the introduction of PHS with respect to the *bau* counterfactual scenario. Fig. 6 (b) presents the annualized overnight investment cost of each PHS configuration, while Fig. 6 (c) presents the resulting net system benefit, calculated as the difference between the system benefit of Fig. 6 (a) and the respective storage fixed cost of Fig. 6 (b) per PHS variant. Fig. 6 (d) depicts the final net system benefit against the *do-minimum* counterfactual scenario, taking into account the value of the firm capacity contribution of each PHS configuration.

It is evident that the introduction of PHS generates a sufficiently high economic benefit to ensure investment viability and still leaves a substantial surplus, even for large PHS configurations. This is the case whether the contribution of storage to capacity adequacy is taken into account or not. Optimal results are obtained for PHS configurations between 700 MW and 1300 MW with a 6-h energy-to-power ratio, when resource adequacy is not valued (*bau* alternative), and between 1250 MW and 1750 MW, again with an energy-to-power ratio of 6-h, when the firm capacity provided by PHS (see Appendix A) is assigned a value (*do minimum* alternative), with a final net system benefit in the order of 75 M€/year.

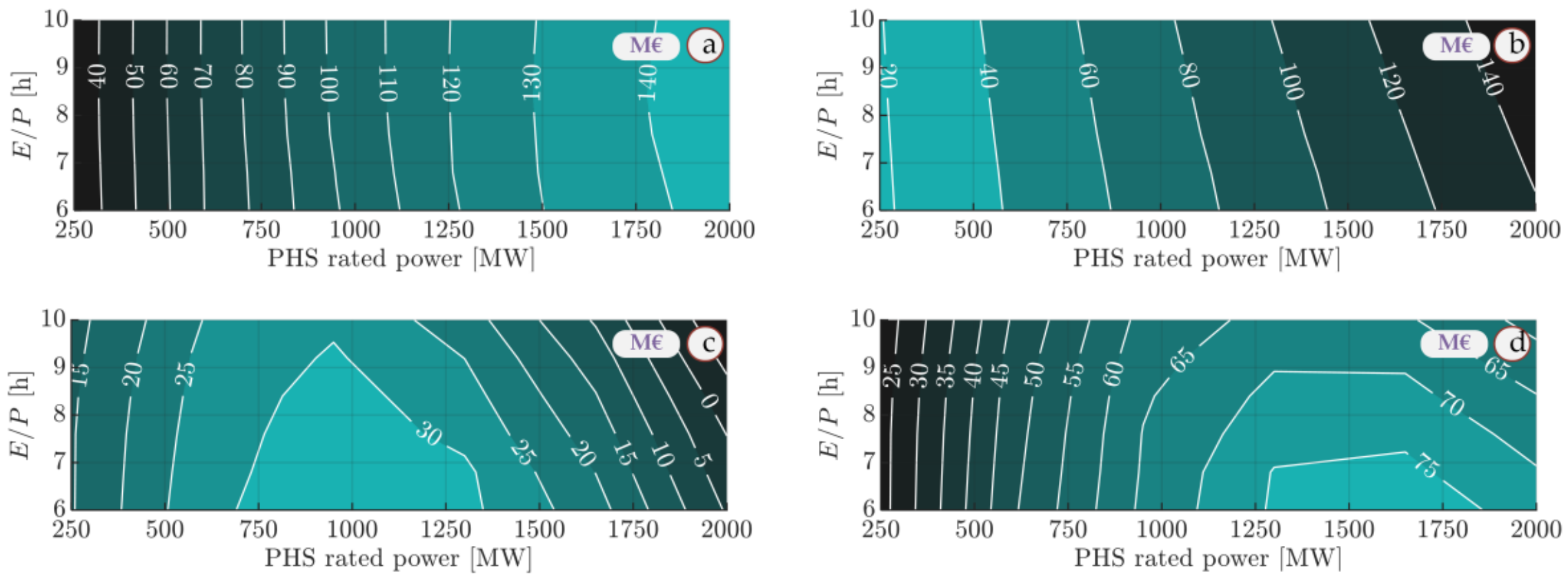

**Fig. 6:** (a) Reduction in annual system operating cost due to the introduction of PHS, (b) PHS annual fixed capital and operating cost, (c) net system benefit against the *bau* counterfactual scenario, and (d) final net system benefit in the *do-minimum* counterfactual scenario. All benefit values expressed in M€.

## 4.2. Benefits through BESS storage

With the introduction of BESS, renewable energy curtailments are significantly reduced, as shown in Fig. 7. RES curtailments reduction for BESS configurations up to 500 MW are independent of storage duration, while at higher capacities the duration starts playing an increasingly important role. This is due to the fact that BESS capacity is predominantly exploited for covering system reserves requirements, allowing the de-commitment of thermal units and thus the increase of RES absorption capability of the system. BESS of a higher power capacity will allow, at the same time, provision of reserves and energy shifting functionality to manage excess RES production of the BESS increases, in which case an increase in energy capacity will also translate in a further reduction of renewable energy curtailments.

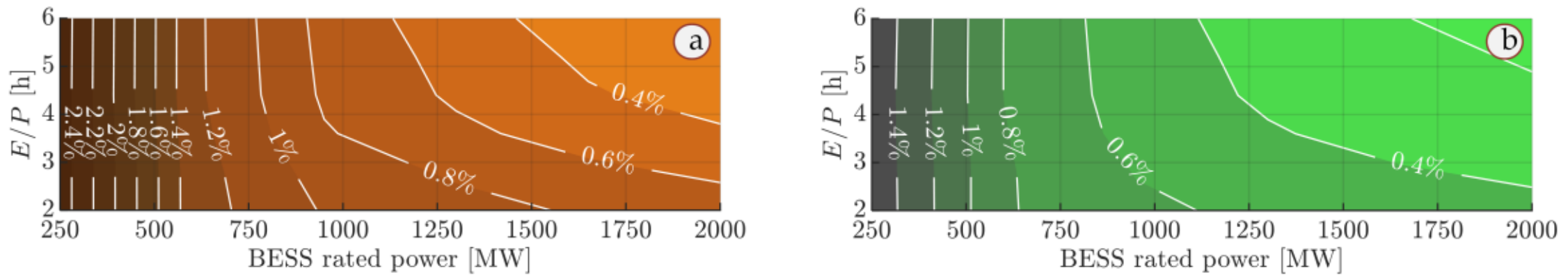

**Fig. 7:** Renewable curtailments in the presence of various BESS configurations as % of the available RES production for (a) solar PV stations and (b) WFs.

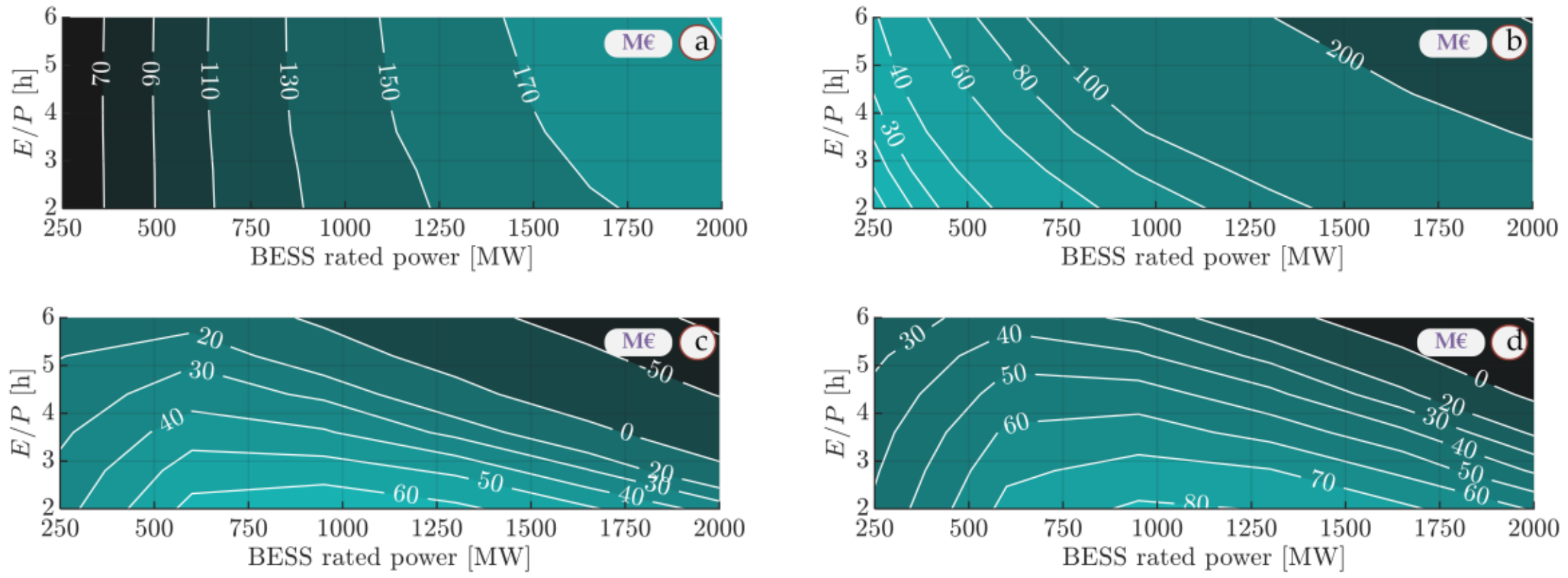

**Fig. 8:** (a) Reduction in annual system operating cost due to the introduction of BESS, (b) annual fixed capital and operating cost, (c) net system benefit against the *bau* counterfactual scenario, and (d) final net benefit in the *do-minimum* counterfactual scenario. All benefit values expressed in M€.

As for PHS, the introduction of BESS will lead to a substantial reduction in the operating cost of the system, as illustrated in Fig. 8 (a) for various configurations. Subtracting the fixed cost of each configuration (Fig. 8 (b)) leads to the eventual net system benefit shown in Fig. 8 (c) and (d) for the *bau* and *do-minimum* counterfactual scenarios. This annual benefit is maximized, reaching ~60 M€ and 80 M€, for 1000 MW BESS of limited energy capacity (2-h), as the increased investment cost of larger configurations cannot be fully compensated by the gains in system operating cost. Notably, this optimum is quite insensitive to the exact sizing of storage, with configurations in the range 500/2-1500/2 in in Fig. 8 (c) and 900/2-1200/2 in Fig. 8 (d) achieving a similar benefit.

## 4.3. A comparison between PHS & BESS

From the analysis performed so far, it is apparent that both PHS or BESS technologies may lead to significant reductions in system cost. To perform a comparative assessment, PHS and BESS configurations of the same power capacity are considered ranging from 250 MW to 2000 MW, with an energy to power ratio of 6-h. The target of this analysis is to evaluate whether batteries are an appropriate substitute of pumped-storage at high duration applications, even if today the norm for BESS systems is up to 4 h.

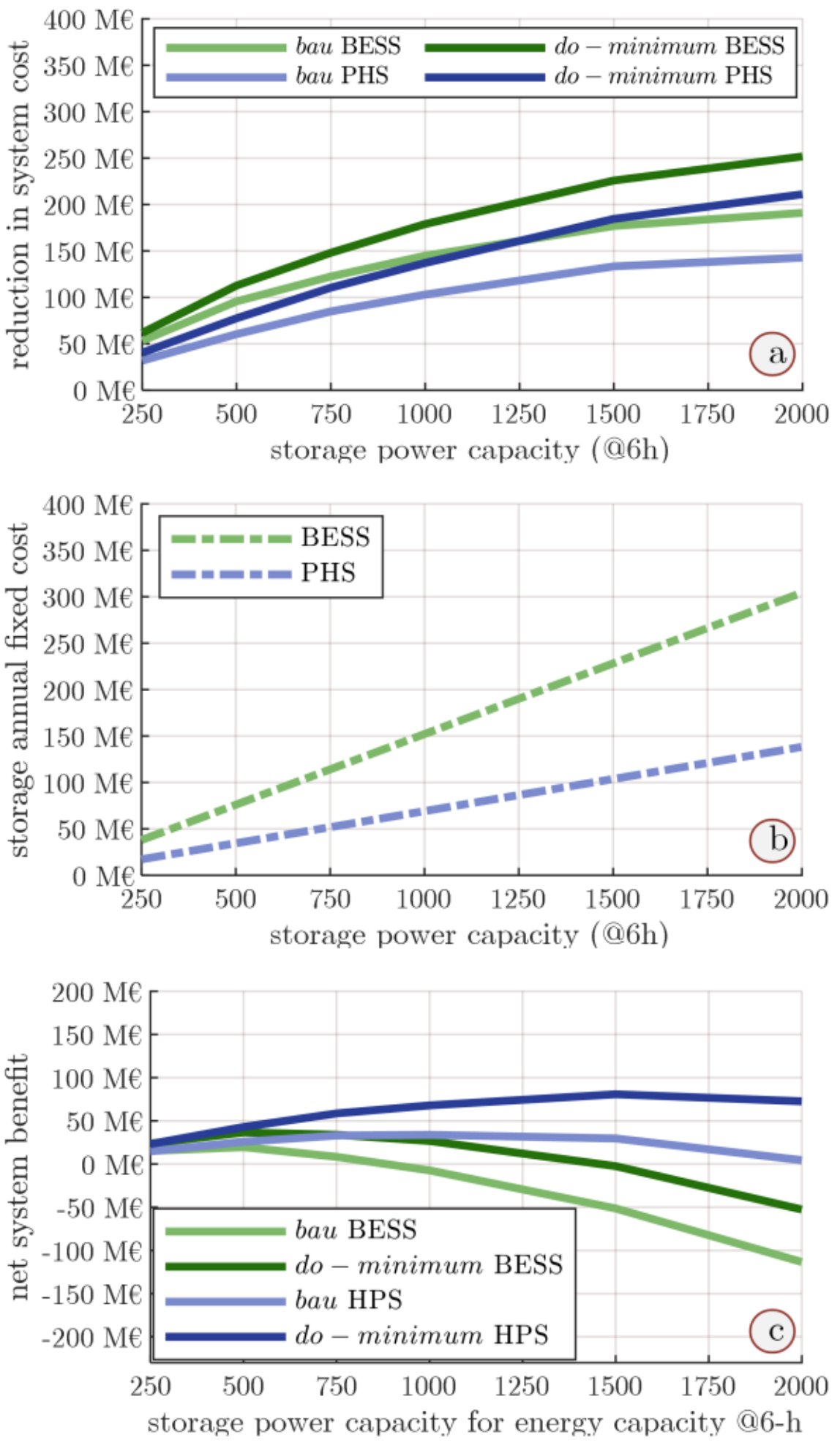

**Fig. 9:** (a) Annual system operation economic benefit due to the introduction of BESS or PHS of the same size, (b) annualized BESS and PHS investment costs, and (c) net system benefit for the *bau* and *do-minimum* counterfactual when either PHS or BESS of the same size are deployed.

The gains in system operating cost obtained through the introduction of BESS or PHS systems of the same 6 h capacity are presented in Fig. 9, along with the respective annualized storage cost and the resulting net benefit for the system. From Fig. 9 (a) it is evident that BESS (green lines) create substantially higher gains compared to PHS (blue lines), by up to 50M€, due to their increased flexibility. Specifically, BESS being inverter-based storage technologies, are practically "online" regardless of their commitment status, they do not have a minimum loading constraint, can transition seamlessly between charging and discharging and they provide fast reserves, with sub-second response times ([62]). On the other hand, BESS investments are still substantially more expensive than PHS stations of the same size (Fig. 9 (b)), especially as regards the cost of energy capacity (Appendix B). When dealing with relatively high durations (6-h or more), their increased cost outweighs the additional economic benefits from their introduction. Hence, in Fig. 9 (c), PHS achieve a higher final net system benefit for any capacity above 500 MW, while for lower sizes the two storage technologies are practically equivalent.

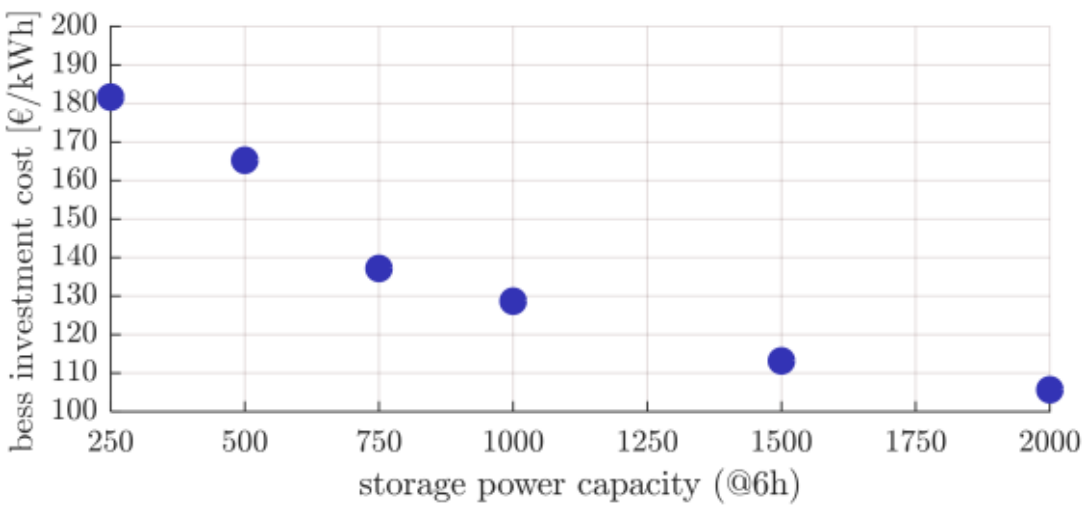

**Fig. 10:** Estimation of required investment costs of 6-h duration BESS, for parity with PHS in terms of resulting system net economic benefit. Note that figures refer to the total investment cost of BESS, expressed in €/kWh, not only the cost of the energy component.

The cost handicap of BESS against PHS is diminishing with the decreasing cost of Li-ion batteries ([63]). Fig. 10 shows the required investment cost of a 6-h BESS, expressed in €/kWh, to achieve the same net system benefit as PHS. It appears that parity is achieved for BESS investment costs in the range of 100-150 €/kWh, substantially lower than today's, but still possible in the long run, after 2035, based on available Li-ion battery cost trend estimates, e.g. [63].

For instance, the initial investment cost for a 1000/6 BESS, expressed in terms of available energy capacity (€/kWh), should be reduced to ~130 €/kWh from the baseline price of ~182 €/kWh that was used to perform the analysis, in order for the BESS to leave the same net economic benefit with a PHS of the same size (1000 MW/ 6-h), which is in the order of 30 M€ for the *bau* counterfactual.

## 4.4. Benefits through PHS & BESS storage portfolios

To evaluate the benefits achieved through the introduction of storage mixtures, 6-h PHS and 2-h BESS storages are selected, given that these capacities were found to be optimal in the previous Sections. The storage mixtures analyzed are presented in Fig. 11. Their aggregate capacity ranges from 1000 to 2000 MW. For each capacity, all technology combinations, from pure PHS to pure BESS, are considered.

Fig. 12 presents the reduction in wind and PV curtailments achieved by the various storage combinations (along with the respective curtailment levels in the absence of any new storage). Similarly, Fig. 13 shows the reduction of $CO_2$ emissions attained due to the presence of storage against the *bau* alternative. RES energy exploitation is optimized for PHS and BESS mixtures, which deliver best results when deployed in combination, as the system simultaneously benefits from the increased flexibility of BESS and the enhanced storage capacity of PHS. $CO_2$ emissions are also reduced when the storage mix comprises BESS and PHS rather than PHS only.

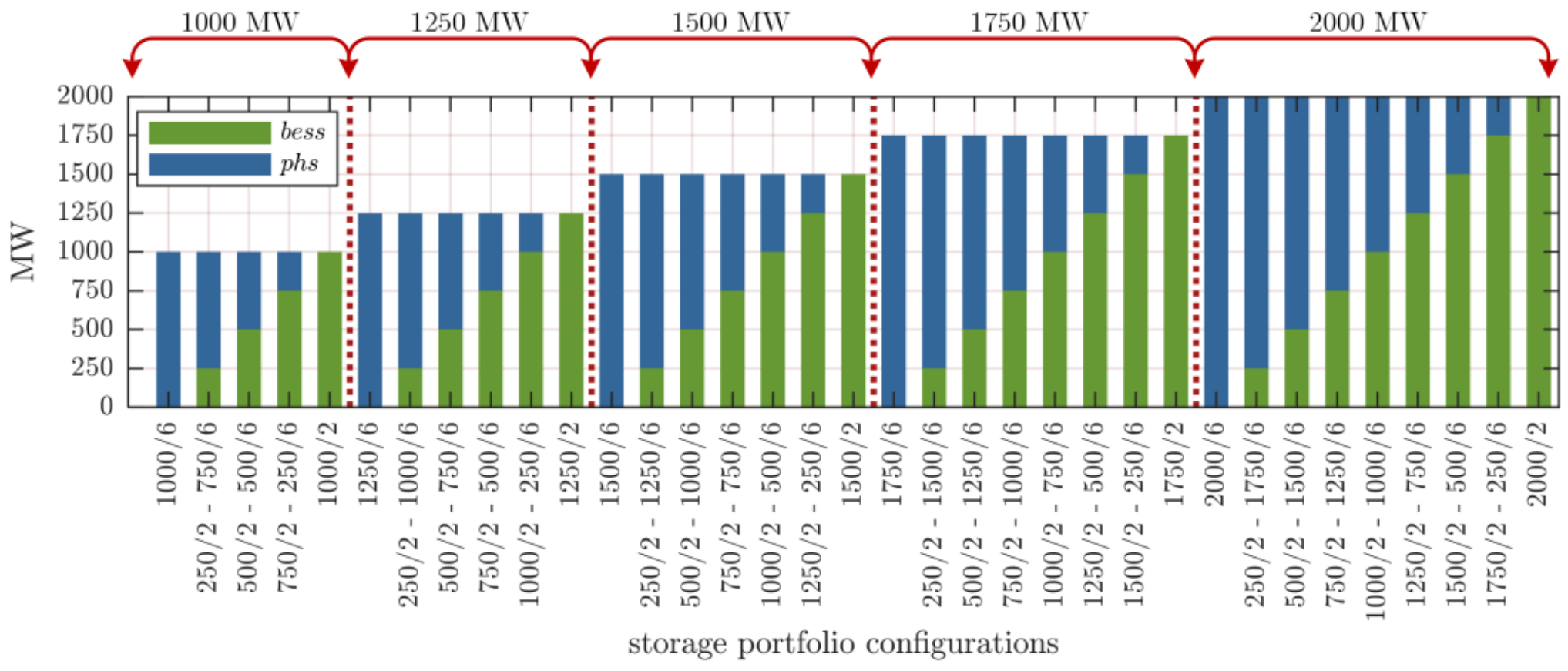

**Fig. 11:** Storage portfolio configurations under examination.

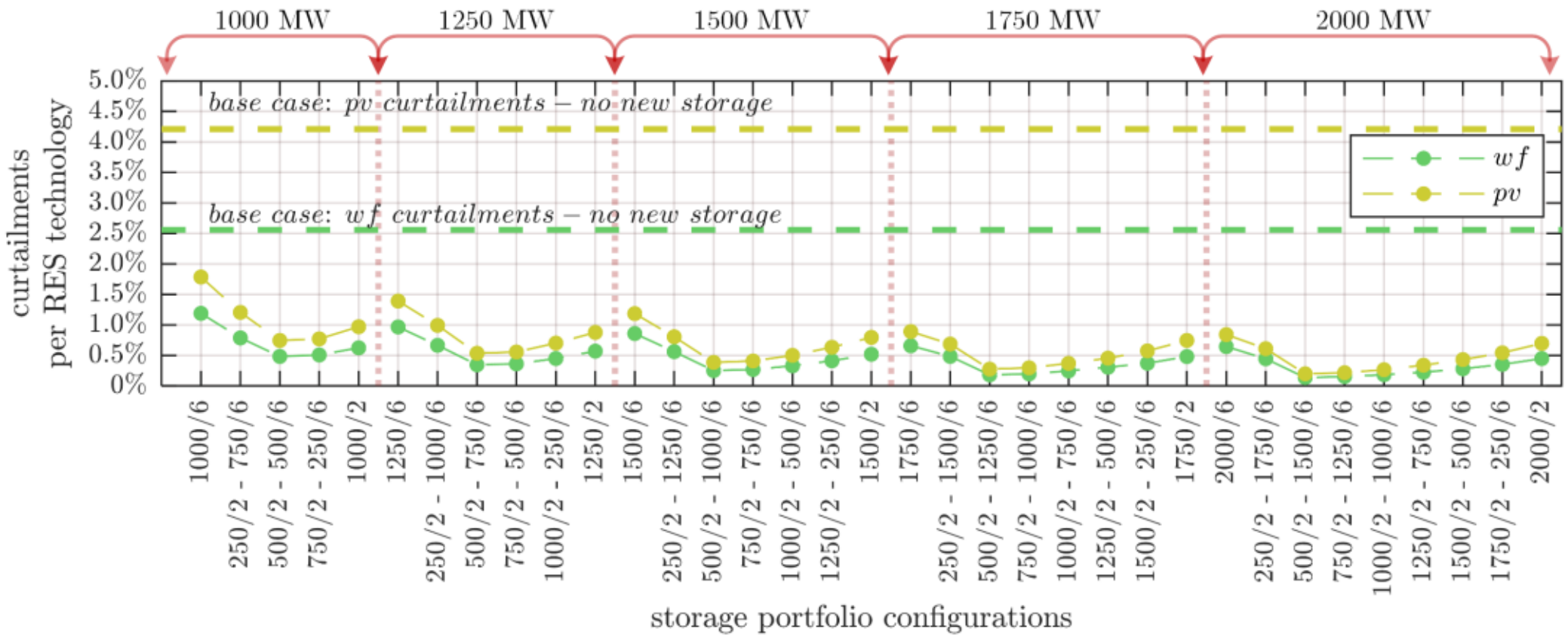

**Fig. 12:** Renewable (PV and WF) energy annual curtailments for the PHS-BESS storage portfolios under evaluation.

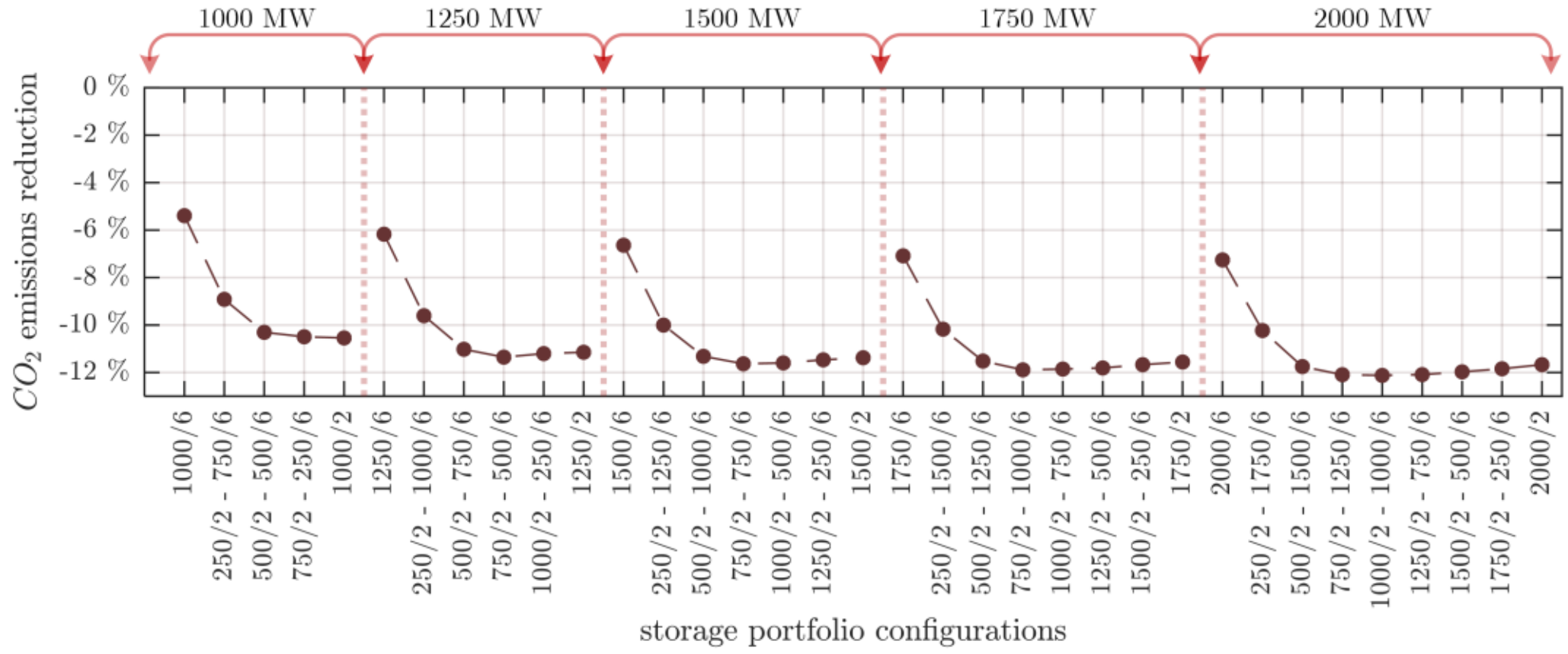

**Fig. 13:** Reduction of $CO_2$ emission for the power system due to the introduction of the PHS-BESS storage portfolios under evaluation.

The combined application of BESS and PHS technologies leads also to enhanced economic benefits for the system, as presented in Fig. 14 for all storage scenarios analyzed. Best results in terms of net system economic benefit are obtained for an aggregate storage capacity between 1000 MW and 1250 MW, when the contribution of storage to capacity adequacy is not considered (the *bau* counterfactual scenario). When the value of the firm capacity provided by storage is considered (the *do-minimum* counterfactual scenario), storages of an aggregate capacity of 1500 MW to 1750 MW yield the best results. Maximum annual benefits reach ~70 M€/y and ~110 M€/y in the *bau* and *do-minimum* cases. The capacity contribution of the storage portfolios is presented in Fig. 15, calculated with the deterministic method of Appendix A.

As regards the composition of storage portfolios, from Fig. 14, it is observed that all optimum storage configurations incorporate BESS of 500 MW/ 2-h in the *do-minimum* case, which may increase up to 750 or 1000 MW in the *bau* case, with PHS providing the complementary storage capacity. Note that this BESS size (500/2) also leads to minimum RES curtailments for all examined storage capacity levels, as illustrated in Fig. 12.

It is important to observe that optimum storage portfolios are loosely defined, in the sense that a wide range of capacities and mixtures will lead to similar levels of economic benefits for the system, while storage capacities well beyond the identified optima still deliver a positive economic equilibrium. Further, it should be emphasized that storage will deliver a range of additional benefits for the power system and the economy in general, beyond the direct impact on generation cost, which are not evaluated in this study and might tip the balance in favor of increased storage capacities, although this remains to be quantified and validated.

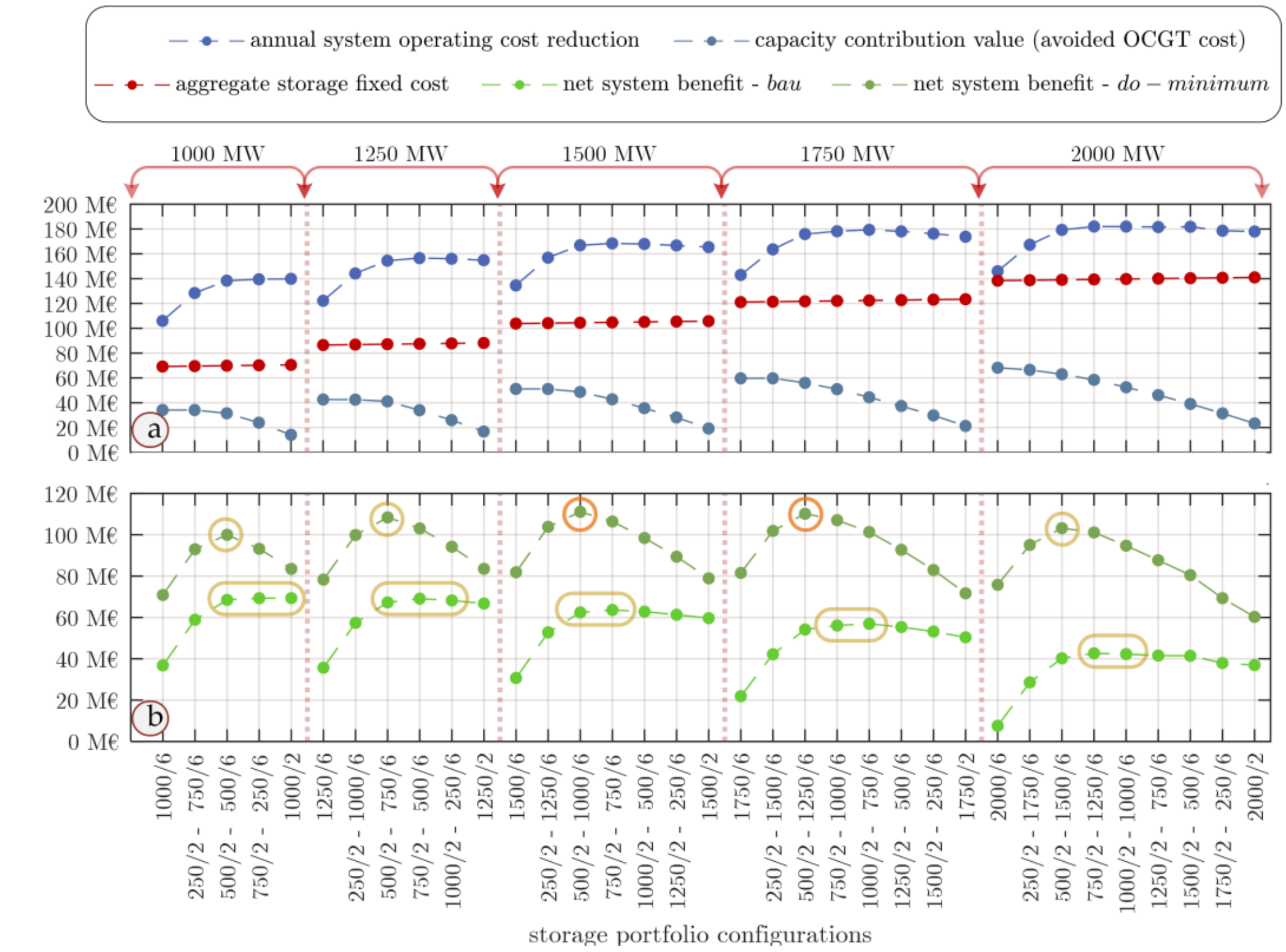

**Fig. 14:** (a) Annual system operating cost reduction, aggregate storage fixed cost and capacity contribution value (avoided OCGT cost); and (b) net system benefit for the *bau* and *do-minimum* counterfactual scenarios, for the storage portfolios under evaluation.

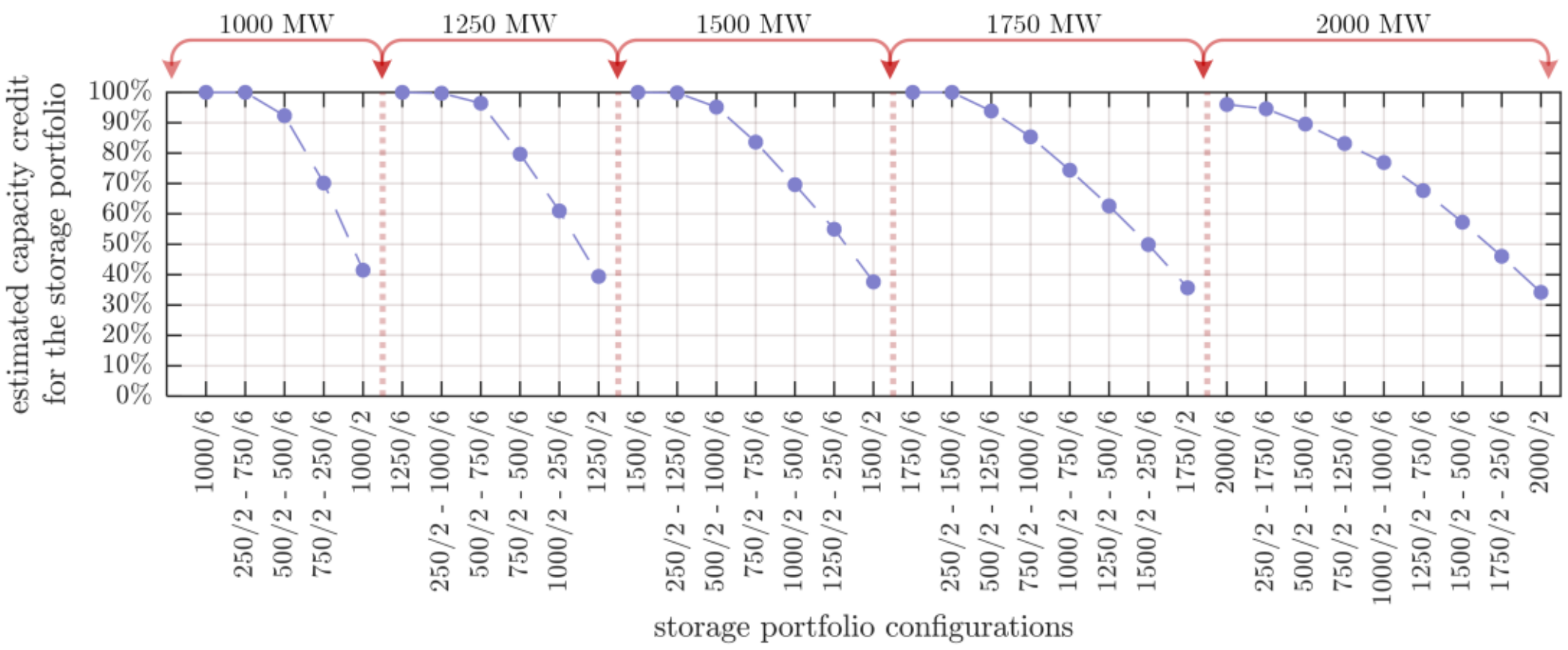

**Fig. 15:** Capacity contribution of the storage portfolios, calculated with the deterministic method of Appendix A.

# 5. Sensitivity analysis

The electricity storage requirements of the system are closely associated with the targeted share of renewables in the energy mix and possibly with the mix of RES technologies selected to achieve the targeted penetration levels. In the following analysis, system storage needs are evaluated for a diversified RES mix, driven by the development of solar PV stations over WFs, as well as for a substantially reduced RES penetration level.

## 5.1. Storage requirements for prioritized PV development

The continuing decline in the cost of PV generation, combined with the increasing obstacles encountered in the development of new wind capacity, especially on-shore, render the hypothesis of a PV-dominated RES generation sector more relevant today than in the past, at least for areas with a high solar potential as the Mediterranean region. The current edition of the Greek NECP for 2030, used as a basis in this analysis, foresees a balanced development with 7.7 GW of PV and 7.0 GW of wind (the "NECP" scenario) to achieve a RES energy penetration level of ca. 60%. In this section, an alternative scenario is evaluated, involving a total capacity of 10.2 GW of PVs and 5.5 GW of WFs (the "PV priority" scenario), to achieve the same overall RES penetration target.

RES curtailments for the new scenario, shown in Fig. 16 (a), indicate that a RES mix dominated by PV generation leads to a considerably more congested system and therefore increased curtailment levels, especially as regards PV stations whose output is highly synchronized. In the absence of new storage capacity, PV curtailments reach ~8%, against 4.5% in the NEPC scenario (Fig. 12 (a)). High curtailments are effectively contained through the introduction of storage, whose capacity needs to be increased by approximately 500 MW in the "PV priority" scenario.

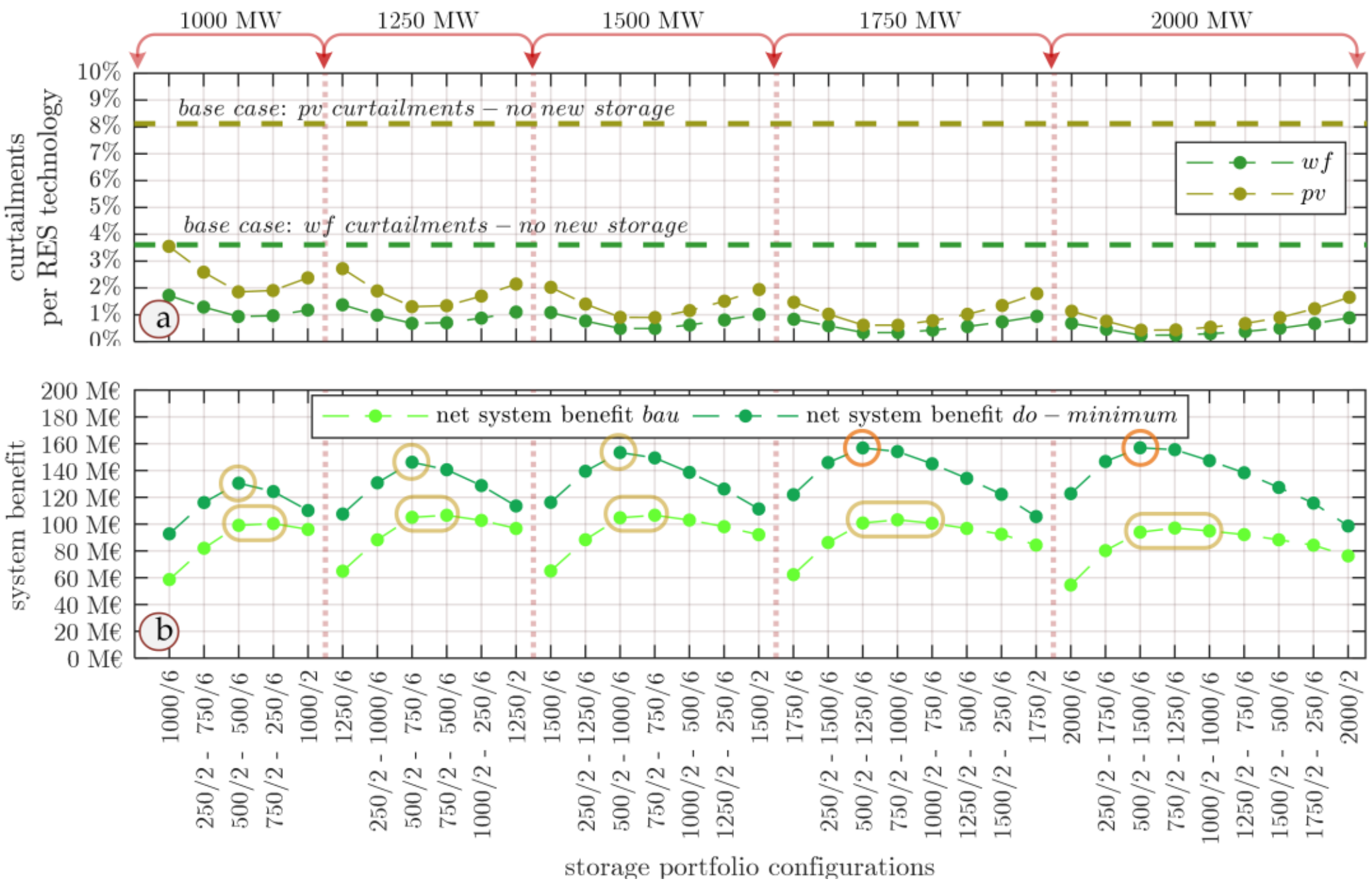

**Fig. 16:** Evaluation of storage portfolios for the "PV priority" scenario. (a) Reduction in renewable energy curtailments and (b) net annual economic benefit for the system.

Economic benefits achieved through the introduction of storage in the "PV priority" scenario, shown in Fig. 16 (b), reach ~100 M€ and ~160 M€ for the *bau* and *do-minimum* cases. Again, storage portfolios are advantageous compared to individual technologies. Compared to the base-case "NECP" scenario, optimum storage capacities now appear substantially increased, ranging from 1250 MW to 1750 MW in the *bau* counterfactual case and exceeding 1750 MW, reaching up to 2000 MW, in the *do-minimum* case. Short-duration BESS storage in all cases represents 500-750 MW of the total storage needs, practically the same as in the "NECP" scenario. This indicates that the capacity of longer-duration PHS storage needs to be enhanced, a fact justified by the increased requirements for energy shifting on the daily load curve due to the concentrated pattern of PV production.

Comparing the gains in the annual system cost achieved in the "PV priority" and base-case "NEPC" scenarios, as illustrated in Fig. 17, it is clear that the benefits from the introduction of storage are substantially higher in the former case, by 38 M€ or 46 M€, comparing the respective optima in the *bau* and *do-minimum* cases, which is primarily attributed to the increased RES curtailments being exploited through storage in the "PV priority" scenario.

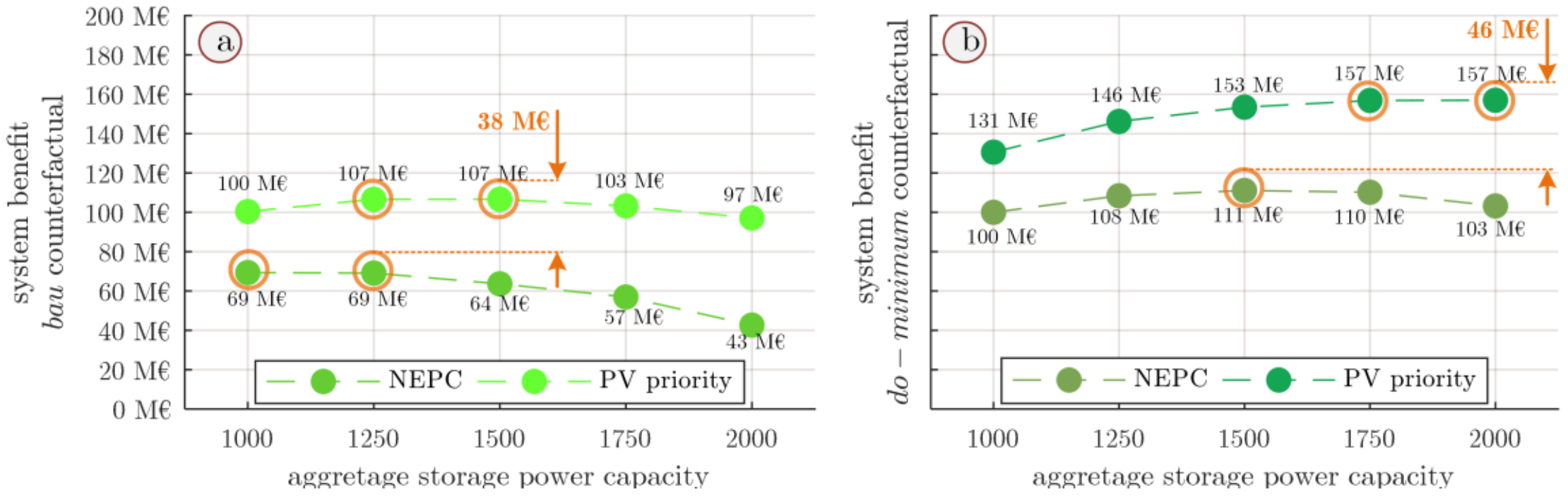

**Fig. 17:** Maximum system benefit at each level of storage portfolio capacity for the "NECP" and "PV priority" cases. (a) *bau* and (b) *do-minimum* counterfactual scenarios.

## 5.2. Storage needs at lower RES penetration rates

To evaluate the needs for storage at lower RES development rates, a scenario is developed with PV and WF capacities reduced to 6.2 GW and 5.5 GW, respectively ("Low RES" scenario), leading to an annual RES penetration rate of ~50%. Otherwise, everything else remains the same as in the base case "NEPC" scenario. Fig. 18 illustrates results obtained for the "Low RES" (in blue) and the "NECP" (in green) scenarios, in the *bau* and *do-minimum* counterfactual cases.

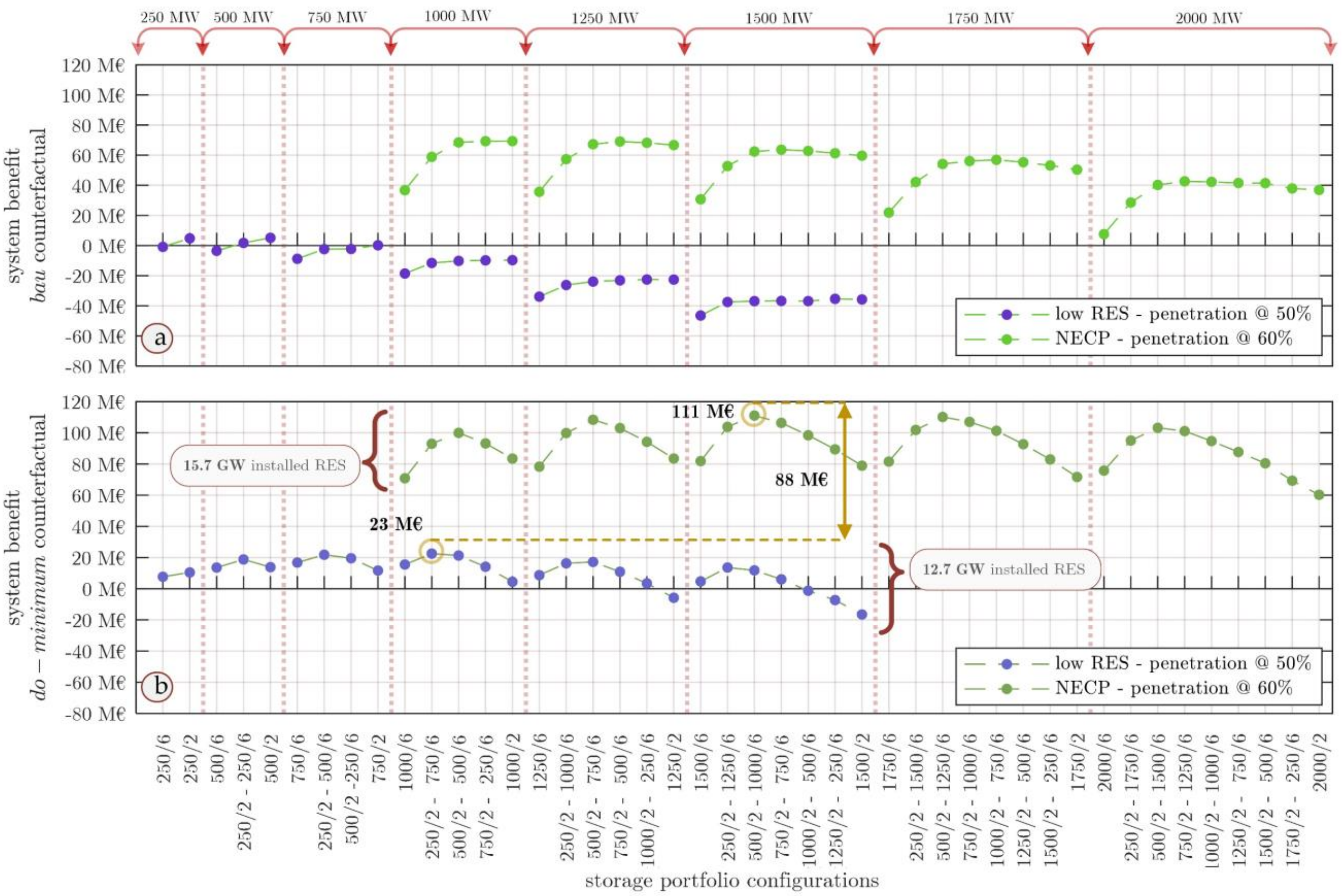

**Fig. 18:** Annual economic benefit for the "NECP" and the "Low RES" scenarios. (a) *bau* and (b) *do-minimum* cases.

From Fig. 18 (a) it is clear that the introduction of new storage facilities at low RES penetration conditions is hardly justified solely on the basis of system operating cost reduction. A marginally positive equilibrium is obtained for BESS configurations 250/2, 500/2 and 750/2, as well as for the combination of BESS 250/2 and PHS 250/6. On the other hand, if capacity adequacy comes into play, as in the *do-minimum* case of Fig. 18(b), the deployment of storages up to 1500 MW can be beneficial for the system, with optimum results obtained for a configuration consisting of BESS 250/2 and PHS 750/6. Still, the net economic benefit does not exceed 20 M€/year, being a fraction of what can be achieved at higher RES penetrations. In this discussion, it should be born in mind that the system already includes 700 MW of open-cycle PHS; therefore, even in the absence of any new facilities, a high-capacity storage basis does exist to serve fundamental needs that will arise even at moderate RES penetrations.

Overall, it is apparent that the introduction of storage is strongly correlated with the level of RES integration in the system. Specific quantitative results derived for the study case interconnected system of the paper indicate that the introduction of storage makes a solid case at RES penetration rates of ca. 50% and above. Nevertheless, such results cannot be safely generalized, as system size, characteristics, mix of RES technologies and interconnectivity with neighboring systems will differentiate one case from another.

## 6. Conclusions

The paper investigates the electricity storage needs to support the integration of an increased share of renewables in an existing power system. The focus is on short- to medium-term solutions, with 2030 being the reference year of analysis, hence the storage technologies of interest are PHS and Li-ion BESS. The analysis adopts a system perspective, aiming to quantify the storage mix that delivers maximum economic benefits for the system. A UC-ED-based methodology with a 24-h optimization horizon has been implemented, which is consecutively executed over the course of a year. Two counterfactual scenarios, without any additional storage deployed in the system, serve as the baseline for the economic evaluation; a *bau* alternative, where it is assumed that the system does not face any resource adequacy issues, and a *do-minimum* approach, where contribution of storage to resource adequacy is assigned a value equal to the avoided fixed cost of OCGT strategic reserve units of an equivalent firm capacity. The Greek power system, in its anticipated development for the year 2030, is used as a study case.

The analysis has highlighted the beneficial impact of storage on reducing the generation cost and mitigating renewable curtailments when targeting annual RES penetration levels of around 60%. Relying on BESS alone, without any new PHS facilities, the optimum capacity is 900 – 1200 MW with a duration of 2-h. Wind and PV energy curtailments would then remain below 0.5% and 1%, respectively. On the other hand, if PHS was the sole storage technology, its optimum capacity would be 700 – 1300 MW, or 1250 – 1750 MW if its contribution to capacity adequacy is taken into account. The optimum duration for PHS assets is 6-h in this particular case study and contemplated RES penetration. RES curtailments would be relatively higher compared to the BESS only case.

Optimal results are obtained through the simultaneous deployment of both BESS and PHS technologies. Specifically, maximum economic benefit was obtained for an aggregate storage capacity in the range of 1250 to 1750 MW, comprising ~500 MW of 2-h BESS systems, with the remaining capacity being PHS of a 6-h duration. This storage mix almost eliminates RES curtailments, while the annual benefit for the system was found to reach 110 M€ compared to the *do-minimum* counterfactual scenario.

Sensitivity analysis regarding the necessity of storage at lower RES penetration levels indicates that new storages, beyond the existing 700 MW of open-loop pumped hydro, starts making economic sense at around 50% annual share of renewables. It was also demonstrated that a mix of renewables driven by solar PVs, leading to the same RES penetration level of ~60%, will impose needs for an increased storage capacity by approx. 500 MW compared to the baseline scenario of balanced wind and solar development.

The findings of this study reveal that the Greek power system, in its transition towards a 60% RES penetration level, from its current 30-35%, will be in need of an enhanced storage portfolio, including both BESS assets of a limited energy capacity and new PHS of higher duration. The Greek authorities have already exploited the results of this study, conducted for the Regulatory Authority of Energy, to establish the right environment to attract and incentivize investments in the appropriate mix of storage. Relevant initiatives have included the development of a new licensing framework for "front-of-the-meter" grid storage and the establishment of investment support mechanisms, in the form of State-aid schemes, to cover the funding gap in case of insufficient market revenues and incentivize investments in storage.

## CRediT authorship contribution statement

**G. N. Psarros:** Conceptualization, Resources, Methodology, Software, Validation, Formal analysis, Investigation, Data curation, Writing - original draft, Visualization. **S. A. Papathanassiou:** Conceptualization, Methodology, Writing - review & editing, Supervision, Project administration, Funding acquisition.

## Declaration of competing interest

The authors declare that they have no known competing financial interests or personal relationships that could have appeared to influence the work reported in this paper.

## Acknowledgments

This work was supported by the Greek Regulatory Authority for Energy.

## **Appendix A:** Contribution of storage to resource adequacy

The contribution of storage to resource adequacy is an open issue, with several different approaches adopted in the literature for the estimation of the capacity credit of energy-constrained storage assets. The most intuitive and simple method is based on the load-leveling technique ([25,52]) to determine the peak reduction potential offered by storage, while more sophisticated methods employ stochastic analysis (the Monte Carlo technique, as in [64]) to include the impact of reliability as well. For the purposes of this paper, the former method is applied to get a rough estimate of the capacity credit of storages of different power and energy capacities. The method, described in [25], is based on the execution of a simplified mixed integer quadratically constrained UC optimization problem.

The capacity contribution of different storage configurations is visualized in Fig A.1. White lines denote derating factors to be applied to each configuration, depending on its energy and power capacity, to determine its contribution in system peak demand mitigation.

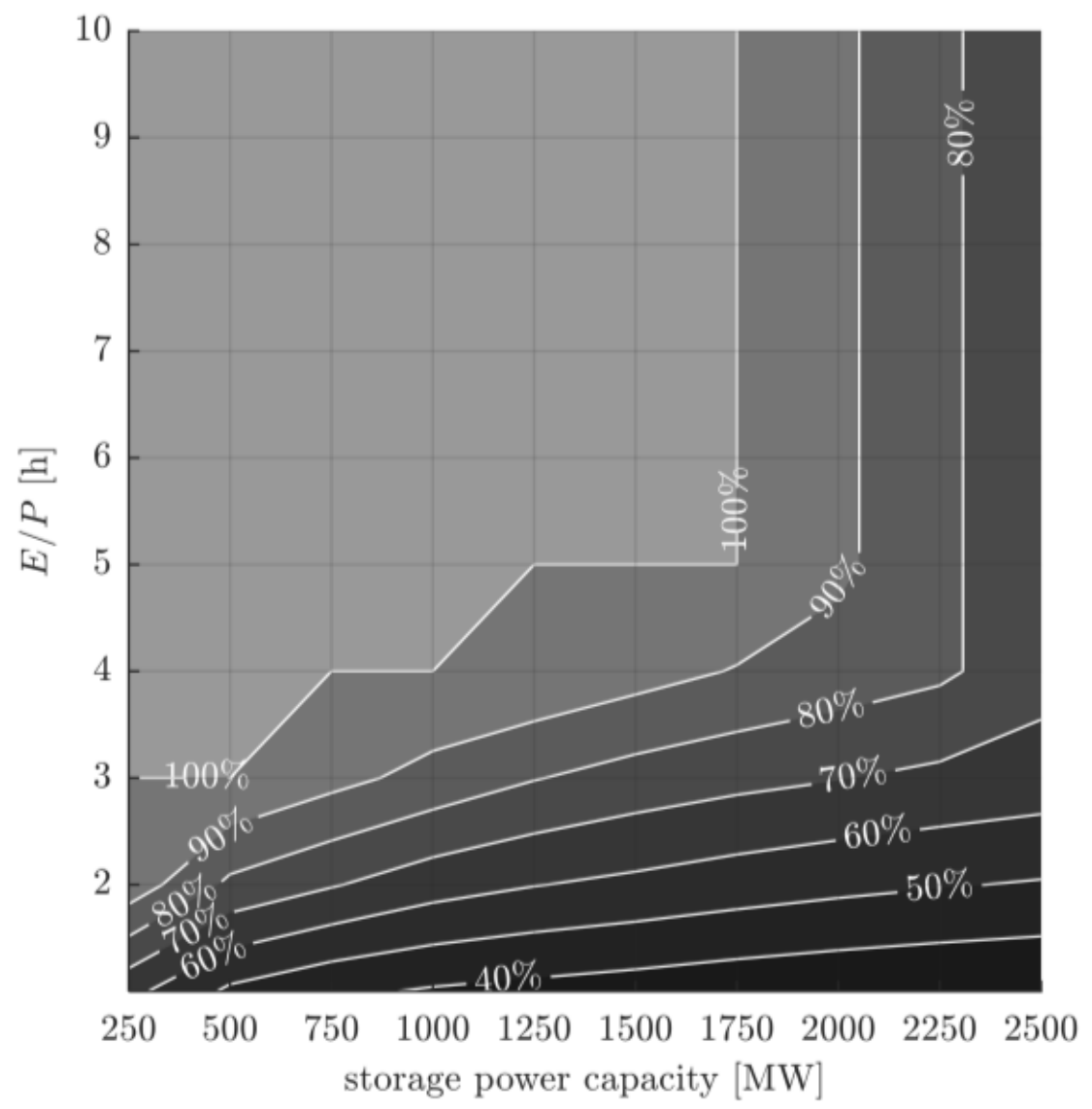

**Fig. A.1:** Estimated capacity credit of different storage configurations based on the load-leveling approach.

# **Appendix B:** Investment cost assumptions, annualized costs & system characteristics

Tables B.1 to B.4 summarize the technical and economic characteristics assumed for the power system of Greece in its near future development. Note that the price assumptions of Table B.4 do not include the impact of disruptive events, such as the effect of the COVID-19 pandemic on the supply chain and the recent energy crisis due to the war in Ukraine, as the study was conducted prior to this situation. It should be noted that the value of storage in support of high RES penetration increases drastically in an environment of high energy prices.

*Table B. 1: Main demand characteristics assumed for the Greek power system.*

| | |
|---|---|
| Annual energy demand | 61.8 TWh |
| *Peak load demand* | 11 GW |
| *Load factor* | 64.13% |

*Table B. 2: Renewables mix assumed for the Greek power system.*

| | |
|---|---|
| Installed WF capacity | 7.0 GW |
| *WF capacity factor* | 28.5% |
| Installed PV capacity | 7.7 GW |
| *PV capacity factor* | 18.1% |
| Other dispatchable RES technologies | 1.0 GW |
| *Run-of-river hydro* | 0.5 GW |
| *Biomass stations* | 0.3 GW |
| *Geothermal stations* | 0.1 GW |
| *Concentrating solar stations* | 0.1 GW |
| **Total RES installed capacity** | **15.7 GW** |

*Table B. 3: Thermal and hydro unit capacities assumed for the Greek power system.*

| | |
|---|---|
| Combined Cycle Gas Turbines | 7.5 GW |
| Open Cycle Gas Turbines | 0.5 GW |
| Hydroelectric plants with reservoirs and natural inflows | 3.5 GW |
| *Without pumping capability* | 2.8 GW |
| *Open loop PHS* | 0.7 GW |
| Demand-side response | 0.6 GW |

| Total capacity | 12.1 GW |
|---|---|

Table B. 4: Variable cost assumptions for thermal units and demand response services.

| | |
|---|---|
| Combined Cycle Gas Turbines* | 65 - 90 €/MWh |
| Open Cycle Gas Turbines* | 120 - 250 €/MWh |
| Demand-side response services | 300 €/MWh |

** includes fuel costs and $CO_2$ emissions rights*

The investment cost assumptions for BESS, PHS and OCGT investments are presented in the following Tables B.5 to B.7. The formula to calculate the annualized overnight cost of an investment is given by (B.1).

$$AnnualizedInvestmentCost = capex\left(\frac{drate}{1-\left(1+drate\right)^{-years}} + opex\right) \tag{B.1}$$

Table B. 5: Economic evaluation parameters for BESS investments.

| Parameter | Value |
|---|---|
| Project lifetime | 25 y |
| Energy capacity component lifetime | 12 y |
| Initial investment cost of energy capacity component (batteries) (*capex*) | 140 €/kWh |
| Cost of energy capacity component (batteries) replacement at 12th year | 100 €/kWh |
| Initial investment cost of power capacity component (*capex*) | 250 €/kW |
| Annual operation and maintenance cost (*opex*), in % of overall project *capex* | 2.5% |
| Discount rate (*drate*) | 8% |

Table B. 6: Economic evaluation parameters for PHS investments.

| Parameter | Value |
|---|---|
| Project lifetime | 40 y |
| Investment cost of energy capacity component (reservoir) (*capex*) | 20 €/kWh |
| Investment cost of power capacity component (*capex*) | 580 €/kW |
| Annual operation and maintenance cost (*opex*), in % of project *capex* | 1.5% |
| Discount rate (*drate*) | 8% |

Table B. 7: Economic evaluation parameters for OGCT investments.

| Parameter | Value |
|---|---|
| Project lifetime | 25 y |
| Investment cost (*capex*) | 300 €/kW |
| Annual operation and maintenance cost (*opex*), in % of project *capex* | 2% |
| Discount rate (*drate*) | 8% |

# **Appendix C:** UC-ED modeling

## Model

The optimization problem is built upon the state-of-the-art MILP technique, incorporating linearized approximations of all non-linear constraints. The mathematical description of the UC-ED problem is briefly presented below.

Specifically, the objective function of the UC-ED and its constituents are presented in constraints (C.1) to (C.8). The objective function of the UC-ED problem, (C.1), consists of the variable operating cost of conventional units (C.2), including fuel costs and the cost of $CO_2$ emissions rights, the variable cost of hydroelectric production (C.3), the cost of imported-exported energy through cross-border interconnections (C.4), the cost of demand response services (C.5), the value of energy stored in the storage facilities of the system at the end of the optimization horizon (C.6), the cost of reserves allocation per reserve type (C.7), and the cost of slack variables (C.8). Equations (C.9) and (C.10) define the energy and reserves equilibria of the power system.

Depending on the type, size, and technical characteristics of the units, some of them may experience increased startup, shut-down, or soak-loading times, which, in turn, may vary based on the time periods that the unit remains offline prior to its startup order. The UC-ED optimization algorithm endogenously determines the operating status of each unit in order to specify its power output. The operating status (online - offline) of each generator and the logical status of commitment are imposed via (C.11) to (C.18), distinguishing between four operational phases: synchronization, soak, dispatchable operation and desynchronization. The fulfillment of the minimum down/up time of thermal generation is ensured by (C.19) and (C.20). The maximum and minimum power output constraints of a generator being in the dispatchable phase bound the unit's operation, considering its ramp-up and ramp-down rates, (C.21). Furthermore, the loading of each online unit operating in the dispatchable phase should account for the allocation of reserves per type and respect the technical maximum and minimum loading of the unit, as imposed by constraints (C.22) and(C.23). The power output of a thermal unit in a specific operating phase is defined by constraints (C.24) to (C.28).

Renewable production is limited by technology availability (C.29). The proportionality of RES curtailments per technology is also accounted for in (C.30) for fair treatment of zero-marginal cost assets by the optimization algorithm.

Storage is managed with the objective of optimizing the overall operation of the power system while adhering to specific technical constraints depending on its technology (PHS or BES). These include the maximum and minimum charging and discharging capability of each facility (constraints (C.31) to (C.33)), the maximum and minimum acceptable levels of stored energy (C.34), the roundtrip efficiency of each storage technology (C.35), as well as the capability of reserves provision per type (constraints (C.36)-(C.37) for BESS and (C.38)-(C.39) for PHS).

Constraints (C.40) to (C.42) concern the operation of hydropower plants with an *upper reservoir* and natural inflows, but *without pumping* facilities. For open-loop pumped-hydro stations, constraints (C.43) to (C.51) apply as well. Demand response services are constrained by (C.52) and (C.53). Cross-border interconnections are modeled based on the net transfer capacity approach via (C.54) to(C.59). Finally, constraints (C.60) and (C.61) delimit reserves allocation to thermal/hydro units according to their technical capability for provision of each reserve type. Constraint (C.61) refers solely to PHS stations.

$$obj = \min\left\{C_{cnv} + C_{hydro} + C_{cr-b} + C_{dr} + C_{es} + C_{aux} + C_{slacks}\right\} \tag{C.1}$$

$$C_{cnv} = \sum_{t,u\in\mathbb{U}_{cnv}} \left( c_u^{ml} \cdot s_{u,t}^{dsp} + \sum_m \left( c_{u,m}^{v} \cdot q_{u,t,m} \right) + \sum_\gamma \left( c_u^{su,\gamma} \cdot y_{u,t}^{\gamma} \right) + c_u^{sd} \cdot z_{u,t} \right) \tag{C.2}$$

$$C_{hydro} = \sum_{t,u\in\mathbb{U}_{hydro},h} c_{u,h}^{o} \cdot \rho_{u,h,t} \tag{C.3}$$

$$C_{cr-b} = \sum_{i,g,t} \left( c_{i,g}^{imp} \cdot q_{i,g,t}^{imp} - c_{i,g}^{exp} \cdot q_{i,g,t}^{exp} \right) \tag{C.4}$$

$$C_{dr} = \sum_t c^{dr} \cdot p_t^{dr} \tag{C.5}$$

$$C_{es} = -\sum_{\xi,t} c_\xi^{stored-e} \cdot \left( \varepsilon_{\xi,T} - \underline{\varepsilon}_\xi \right) \tag{C.6}$$

$$C_{aux} = \sum_{u\in\mathbb{U},t,e} c_u^{e\pm} \cdot r_{u,t}^{e\pm} + \sum_{\xi,t,e} c_\xi^{e\pm} \cdot r_{\xi,t}^{e\pm}, \qquad e = \{fcr, a-frr, m-frr\} \tag{C.7}$$

$$C_{slacks} = \sum_t \left(\varphi^{ens} \cdot p_t^{ens} + \varphi^{e\pm} \cdot sl_t^{e\pm}\right) \tag{C.8}$$

$$\sum_{u\in\mathbb{U}} p_{u,t} + \sum_{u\in\mathbb{U}_{ohps}} p_{u,t}^{ohps-d} + \sum_\xi p_{\xi,t}^{ess-d} + \sum_i p_{i,t}^{imp} + \sum_{res} p_{res,t} + p_t^{dr} + p_t^{ens} = p_t^d + \sum_{u\in\mathbb{U}_{ohps}} p_{u,t}^{ohps-c} + \sum_\xi p_{\xi,t}^{ess-c} + \sum_i p_{i,t}^{exp} \tag{C.9}$$

$$\sum_{u\in\mathbb{U}} r_{u,t}^{e\pm} + \sum_\xi r_{\xi,t}^{e\pm} + sl_t^{e\pm} = rr_t^{e\pm}, \qquad e = \{fcr, a-frr, m-frr\} \tag{C.10}$$

$$y_{u,t} + z_{u,t} \le 1 \tag{C.11}$$

$$y_{u,t} - z_{u,t} \ge s_{u,t} - s_{u,t-1} \tag{C.12}$$

$$s_{u,t} = s_{u,t}^{syn} + s_{u,t}^{soak} + s_{u,t}^{dsp} + s_{u,t}^{dsyn} \tag{C.13}$$

$$y_{u,t}^\gamma \le \sum_{k=t-T_{u,\gamma}+1}^{t-T_{u,\gamma-1}} z_{u,k} \tag{C.14}$$

$$\sum_\gamma y_{u,t}^\gamma = y_{u,t} \tag{C.15}$$

$$s_{u,t}^{syn} = \sum_\gamma \sum_{k=t-T_{u,\gamma}^{syn}+1}^{t} y_{u,t}^\gamma \tag{C.16}$$

$$s_{u,t}^{soak} = \sum_\gamma \sum_{k=t-T_{u,\gamma}^{syn}-T_{u,\gamma}^{soak}+1}^{t-T_{u,\gamma}^{syn}} y_{u,t}^\gamma \tag{C.17}$$

$$s_{u,t}^{des} = \sum_{k=t+1}^{t+T_u^{des}-1} z_{u,k} \tag{C.18}$$

$$\sum_{k=t-T_u^{stop}+1}^{t} z_{u,k} \le 1 - s_{u,t} \tag{C.19}$$

$$\sum_{k=t-\left(T_{u,\gamma}^{syn}+T_{u,\gamma}^{soak}+T_{u,t}^{run}+T_u^{des}\right)+1}^{t} y_{u,t}^\gamma \le s_{u,t} \tag{C.20}$$

$$-\upsilon_u \cdot s_{u,t}^{dsp} - \overline{p}_u \cdot \left(s_{u,t}^{syn} + s_{u,t}^{soak}\right) \le p_{u,t-1} - p_{u,t} \le d_u \cdot s_{u,t}^{dsp} + \overline{p}_u \cdot \left(z_{u,t} + s_{u,t}^{des}\right) \tag{C.21}$$

$$p_{u,t} + \sum_e r_{u,t}^{e+} \le p_{u,t}^{soak} + p_{u,t}^{des} + \overline{p}_u \cdot s_{u,t}^{dsp} \tag{C.22}$$

$$p_{u,t} - \sum_e r_{u,t}^{e-} \ge p_{u,t}^{soak} + p_{u,t}^{des} + \underline{p}_u \cdot s_{u,t}^{dsp} \tag{C.23}$$

$$p_{u,t}^{dsp} = \underline{p}_u + \sum_m q_{u,t,m}, \quad \forall u \in \mathbb{U}_{cnv} \tag{C.24}$$

$$0 \le q_{u,t,m} \le \overline{q}_{u,t,m}, \quad \forall u \in \mathbb{U}_{cnv} \tag{C.25}$$

$$p_{u,t}^{syn} = 0 \tag{C.26}$$

$$p_{u,t}^{soak} = \sum_\gamma \sum_{k=t-T_{u,\gamma}^{syn}-T_{u,\gamma}^{soak}+1}^{t-T_{u,\gamma}^{syn}} \left\{ y_{u,t}^\gamma \cdot p_u^{sn} + y_{u,t}^\gamma \cdot \left(t - T_{u,\gamma}^{syn} - k\right) \cdot \frac{\underline{p}_u - p_u^{sn}}{T_{u,\gamma}^{soak}} \right\} \tag{C.27}$$

$$p_{u,t}^{des} = \sum_{k=t+1}^{t+T_u^{des}} z_{u,k} \cdot \frac{\underline{p}_u}{T_u^{des}} \cdot \left(k - t\right) \tag{C.28}$$

$$p_{res,t} + x_{res,t} = p_{res,t}^a \tag{C.29}$$

$$\frac{x_{w,t}}{p_{w,t}^a} = \frac{x_{pv,t}}{p_{pv,t}^a} = \frac{x_{other,t}}{p_{other,t}^a} \tag{C.30}$$

$$\underline{p}_\xi^{ess-d} \cdot s_{\xi,t}^d \le p_{\xi,t}^{ess-d} \le \overline{p}_\xi^{ess-d} \cdot s_{\xi,t}^d \tag{C.31}$$

$$\underline{p}_\xi^{ess-c} \cdot s_{\xi,t}^c \le p_{\xi,t}^{ess-c} \le \overline{p}_\xi^{ess-c} \cdot s_{\xi,t}^c \tag{C.32}$$

$$s^{c}_{\xi,t} + s^{d}_{\xi,t} \leq 1 \tag{C.33}$$

$$\underline{\varepsilon}_{\xi} \leq \varepsilon_{\xi,t} \leq \overline{\varepsilon}_{\xi} \tag{C.34}$$

$$\varepsilon_{\xi,t} = \varepsilon_{\xi,t-1} + p^{ess-c}_{\xi,t} \cdot \sqrt{n_{\xi}} - p^{ess-d}_{\xi,t} \Big/ \sqrt{n_{\xi}} \tag{C.35}$$

$$p^{ess-d}_{\xi,t} + \sum_{e} r^{e+}_{\xi,t} \leq \overline{p}^{ess-d}_{\xi} + p^{ess-c}_{\xi,t}, \quad \forall \xi \in \mathbb{X}_{bess} \tag{C.36}$$

$$p^{ess-c}_{\xi,t} + \sum_{e} r^{e-}_{\xi,t} \leq \overline{p}^{ess-c}_{\xi} + p^{ess-d}_{\xi,t}, \quad \forall \xi \in \mathbb{X}_{bess} \tag{C.37}$$

$$p_{\xi,t} + \sum_{e} r^{e+}_{\xi,t} \leq \overline{p}^{ess-d}_{\xi} \cdot s^{d}_{\xi,t}, \quad \forall \xi \in \mathbb{X}_{hps} \tag{C.38}$$

$$p_{\xi,t} - \sum_{e} r^{e-}_{\xi,t} \geq \underline{p}^{ess-d}_{\xi} \cdot s^{d}_{\xi,t}, \quad \forall \xi \in \mathbb{X}_{hps} \tag{C.39}$$

$$p_{u,t} = \sum_{h} \rho_{u,h,t} + \omega_{u,t}, \quad \forall u \in \mathbb{U}_{hydro} \tag{C.40}$$

$$\sum_{t} \omega_{u,t} \leq \ell_{u}, \quad \forall u \in \mathbb{U}_{hydro} \tag{C.41}$$

$$\sum_{t} \rho_{u,h,t} \leq j_{u,h}, \quad \forall u \in \mathbb{U}_{hydro} \tag{C.42}$$

$$\underline{p}_{u} \cdot s^{ohps-d}_{u,t} \leq p^{ohps-d}_{u,t} \leq \overline{p}_{u} \cdot s^{ohps-d}_{u,t}, \quad \forall u \in \mathbb{U}^{ohps}_{hydro} \tag{C.43}$$

$$\underline{p}^{ohps-c}_{u} \cdot s^{ohps-c}_{u,t} \leq p^{ohps-c}_{u,t} \leq \overline{p}^{ohps-c}_{u} \cdot s^{ohps-c}_{u,t}, \quad \forall u \in \mathbb{U}^{ohps}_{hydro} \tag{C.44}$$

$$s^{ohps-c}_{u,t} + s^{ohps-d}_{u,t} \leq 1, \quad \forall u \in \mathbb{U}^{ohps}_{hydro} \tag{C.45}$$

$$s^{ohps-c}_{u,t} + s_{u,t} \leq 1, \quad \forall u \in \mathbb{U}^{ohps}_{hydro} \tag{C.46}$$

$$s^{ohps-d}_{u,t} \leq s^{dsp}_{u,t}, \quad \forall u \in \mathbb{U}^{ohps}_{hydro} \tag{C.47}$$

$$\varepsilon^{ohps}_{u,t} = \varepsilon^{ohps}_{u,t-1} + p^{ohps-c}_{u,t} \cdot \sqrt{n^{ohps}_{u}} - p^{ohps-d}_{u,t} \Big/ \sqrt{n^{ohps}_{u}}, \quad \forall u \in \mathbb{U}^{ohps}_{hydro} \tag{C.48}$$

$$\underline{\varepsilon}^{ohps}_{u,t} \leq \varepsilon^{ohps}_{u,t} \leq \overline{\varepsilon}^{ohps}_{u,t}, \quad \forall u \in \mathbb{U}^{ohps}_{hydro} \tag{C.49}$$

$$p^{dsp}_{u,t} + p^{ohps-d}_{u,t} + \sum_{e} r^{e+}_{u,t} \leq \overline{p}_{u} \cdot s^{dsp}_{u,t}, \quad \forall u \in \mathbb{U}^{ohps}_{hydro} \tag{C.50}$$

$$p^{dsp}_{u,t} + p^{ohps-d}_{u,t} - \sum_{e} r^{e-}_{u,t} \geq \underline{p}_{u} \cdot s^{dsp}_{u,t}, \quad \forall u \in \mathbb{U}^{ohps}_{hydro} \tag{C.51}$$

$$p^{dr}_{t} \leq \overline{p}^{dr}_{t} \tag{C.52}$$

$$\sum_{t} p^{dr}_{t} \leq dd \tag{C.53}$$

$$p^{imp}_{i,t} - p^{exp}_{i,t} \leq ntc^{imp}_{i,t} \tag{C.54}$$

$$p^{exp}_{i,t} - p^{imp}_{i,t} \leq ntc^{exp}_{i,t} \tag{C.55}$$

$$p^{imp}_{i,t} = \sum_{g} q^{imp}_{i,g,t} \tag{C.56}$$

$$p^{exp}_{i,t} = \sum_{g} q^{exp}_{i,g,t} \tag{C.57}$$

$$0 \leq q^{imp}_{i,g,t} \leq \overline{q}^{imp}_{i,g,t} \tag{C.58}$$

$$0 \leq q^{exp}_{i,g,t} \leq \overline{q}^{exp}_{i,g,t} \tag{C.59}$$

$$r^{e\pm}_{u,t} \leq rc^{e\pm}_{u} \cdot s_{u,t}, \qquad e = \{fcr, a-frr, m-frr\} \tag{C.60}$$

$$r^{e\pm}_{\xi,t} \leq rc^{e\pm}_{\xi} \cdot s^{d}_{\xi,t}, \qquad e = \{a-frr, m-frr\} \cup \xi \in \mathbb{X}_{hps} \tag{C.61}$$

## Notation

### Sets

$\mathbb{E}$ Set of indices of reserves types

$\mathbb{G}$ Set of indices for the power offer blocks from neighboring countries

| | |
|---|---|
| $\mathbb{H}$ | Set of indices for the energy offers from hydroelectric power stations |
| $\mathbb{I}$ | Set of indices of cross border interconnections |
| $\mathbb{M}$ | Set of indices for the blocks of the variable operating cost function of each conventional unit |
| $\mathbb{R}$ | Set of indices of the renewable energy sources |
| $\mathbb{T}$ | Set of indices of time intervals within optimization horizon |
| $\mathbb{U}$ | Set of indices of dispatchable units |
| $\mathbb{X}$ | Set of indices of flexible standalone storage units |
| $\mathbb{Y}$ | Set of indices of unit startup types |
| $\mathbb{U}_{cnv} \subset \mathbb{U}$ | Subset of indices of conventional units |
| $\mathbb{U}_{hydro} \subset \mathbb{U}$ | Subset of indices of hydroelectric units |
| $\mathbb{U}_{hydro}^{ohps} \subset \mathbb{U}_{hydro}$ | Subset of indices of open-loop hydro pumped storage units |
| $\mathbb{X}_{hps} \subset \mathbb{X}$ | Subset of indices of hydro-pumped storage stations |
| $\mathbb{X}_{bess} \subset \mathbb{X}$ | Subset of indices of battery energy storage stations |

**Indices**

| | |
|---|---|
| $e \in \mathbb{E}$ | Index of reserves type (*FCR/a-FRR/m-FRR*) |
| $g \in \mathbb{G}$ | Index of power offer blocks from neighboring countries |
| $h \in \mathbb{H}$ | Index of energy offers from hydroelectric power stations |
| $i \in \mathbb{I}$ | Index of cross border interconnections |
| $m \in \mathbb{M}$ | Index of blocks of the linearized cost function of each conventional unit |
| $res \in \mathbb{R}$ | Index of renewable energy sources types (*wind farms/solar PVs/other RES*) |
| $t, k \in \mathbb{T}$ | Index of time intervals of the optimization horizon |
| $u \in \mathbb{U}$ | Index of dispatchable generating units |
| $\xi \in \mathbb{X}$ | Index of standalone storage units (*PHS, BESS*) |
| $\gamma \in \mathbb{Y}$ | Index of unit start-up types (*cold/warm/hot*) |

**Binary variables**

| | |
|---|---|
| $s_{u,t}$ | Binary variable equal to 1 if unit *u* is online at *t* |
| $s_{u,t}^{dsp}$ | Binary variable equal to 1 if unit *u* is at dispatchable phase at *t* |
| $s_{u,t}^{dsyn}$ | Binary variable equal to 1 if unit *u* is at desynchronization phase at *t* |
| $s_{u,t}^{soak}$ | Binary variable equal to 1 if unit *u* is at soak phase at *t* |
| $s_{u,t}^{syn}$ | Binary variable equal to 1 if unit *u* is at synchronization phase at *t* |
| $s_{\xi,t}^{c}$ | Binary variable equal to 1 if storage $\xi$ charges at *t* |
| $s_{\xi,t}^{d}$ | Binary variable equal to 1 if storage $\xi$ discharges at *t* |
| $s_{u,t}^{ohps-c}$ | Binary variable equal to 1 if oHPS unit *u* discharges at *t* |
| $s_{u,t}^{ohps-d}$ | Binary variable equal to 1 if oHPS unit *u* discharges at *t* |
| $y_{u,t}^{\gamma}$ | Binary variable equal to 1 if a startup of type $\gamma$ for unit *u* at *t* is selected |
| $y_{u,t}$ | Binary variable equal to 1 if unit *u* starts up at *t* |
| $z_{u,t}$ | Binary variable equal to 1 if unit *u* shuts down at *t* |

**Continuous variables**

| | |
|---|---|
| $C_{cnv}$ | Variable operating cost of conventional units over the optimization horizon |
| $C_{hydro}$ | Cost attributed to hydroelectric production over the optimization horizon |
| $C_{cr-b}$ | Cost attributed to energy transactions over the cross border interconnections over the optimization horizon |
| $C_{dr}$ | Cost of demand response services over the optimization horizon |
| $C_{es}$ | Value of energy stored in the electric storages at the end of the optimization horizon |
| $C_{aux}$ | Cost of allocated reserves over the optimization horizon |
| $C_{slacks}$ | Cost of slack variables over the optimization horizon |
| $p_{u,t}$ | Production level of unit *u* at *t* |
| $p_{u,t}^{dsp}$ | Production level of unit *u* at *t* when in dispatchable phase |
| $p_{u,t}^{syn}$ | Production level of unit *u* at *t* when in synchronization phase |
| $p_{u,t}^{soak}$ | Production level of unit *u* at *t* when in soak phase |
| $p_{u,t}^{des}$ | Production level of unit *u* at *t* when in desynchronization phase |
| $p_{t}^{dr}$ | Activated demand response services at *t* |
| $p_{u,t}^{ohps-c}$ | Charging level of oHPS *u* at *t* |
| $p_{u,t}^{ohps-d}$ | Discharging level of oHPS *u* at *t* |
| $p_{\xi,t}^{ess-c}$ | Charging level of storage $\xi$ at *t* |
| $p_{\xi,t}^{ess-d}$ | Discharging level of storage $\xi$ at *t* |
| $p_{i,t}^{exp}$ | Total exports levels from border *i* at *t* |
| $p_{i,t}^{imp}$ | Total imports levels from border *i* at *t* |
| $p_{res,t}$ | Production level of RES type *res* at *t* |
| $p_{t}^{ens}$ | Energy not served at *t* |
| $q_{i,g,t}^{imp}$ | Imports level of border *i* at block *g* at *t* |
| $q_{i,g,t}^{exp}$ | Exports level of border *i* at block *g* at *t* |
| $q_{u,t,m}$ | Production level of conventional unit *u* at block *m* at *t* |
| $r_{u,t}^{e\pm}$ | Allocated upwards(+)/downwards(-) reserves to *u* at *t* |
| $r_{\xi,t}^{e\pm}$ | Allocated upwards(+)/downwards(-) reserves to $\xi$ at *t* |
| $sl_{t}^{e\pm}$ | Activated slack variable for reserves requirements *e* constraints at *t* |
| $x_{res,t}$ | Renewable curtailments of RES type *res* |
| $\varepsilon_{\xi,t}$ | Stored energy levels of $\xi$ at *t* |
| $\varepsilon_{u,t}^{ohps}$ | Stored energy levels of open-loop HPS *u* at *t* |
| $\rho_{u,h,t}$ | Dispatched hydro energy over mandatory injections, of block *h* for hydroelectric unit *u* at *t* |
| $\omega_{u,t}$ | Dispatched hydro mandatory injections for hydroelectric unit *u* at *t* |
| **Parameters** | |
| $c_{u}^{ml}$ | Cost of operating at minimum loading for unit *u* |
| $c_{u,m}^{v}$ | Slope (marginal cost) of each block *m* of conventional unit *u* |
| $c_{u}^{su,\gamma}$ | Startup cost of unit *u* for startup type $\gamma$ |
| $c_{u}^{sd}$ | Shut-down cost of unit *u* |
| $c_{u,h}^{o}$ | Cost of energy offer *h* of hydroelectric unit *u* |
| $c_{i,g}^{imp}$ | Imports cost for each block *g* of border *i* |
| $c_{i,g}^{exp}$ | Exports cost for each block *g* of border *i* |
| $c^{dr}$ | Cost of demand response services |
| $c_{\xi}^{stored-e}$ | Value of stored energy in storage $\xi$ at the end of optimization horizon |
| $c_{u}^{e\pm}$ | Virtual cost attributed to allocated reserves of type *e* and unit *u* |
| $c_{\xi}^{e\pm}$ | Virtual cost attributed to allocated reserves of type *e* and storage $\xi$ |
| $d_{u}$ | Ramp down rate of unit *u* |
| $dd$ | Maximum level of activated demand response services within the optimization horizon |
| $j_{u,h}$ | Available hydro energy over mandatory of block *h* to be dispatched for hydroelectric unit *u* within the optimization horizon |
| $\ell_{u}$ | Mandatory hydro injections for hydroelectric unit *u* within the optimization horizon |
| $n_{\xi}$ | Roundtrip efficiency of storage $\xi$ |
| $n_{u}^{ohps}$ | Roundtrip efficiency of open-loop HPS station *u* |
| $ntc_{i,t}^{imp}$ | Net transfer import capacity of border *i* at *t* |
| $ntc_{i,t}^{exp}$ | Net transfer export capacity of border *i* at *t* |
| $p_{t}^{d}$ | Load demand at *t* |
| $p_{res,t}^{a}$ | Available production of RES type *res* at *t* |
| $p_{u}^{sn}$ | Synchronization load of unit *u* |
| $\overline{p}_{t}^{dr}$ | Maximum level of demand response services at *t* |
| $\overline{p}_{u}$ | Maximum power output of unit *u* |
| $\underline{p}_{u}$ | Minimum power output (minimum loading) of unit *u* |
| $\overline{p}_{\xi}^{ess-d}$ | Maximum discharging capacity of storage $\xi$ |
| $\underline{p}_{\xi}^{ess-d}$ | Minimum discharging capacity of storage $\xi$ |
| $\overline{p}_{\xi}^{ess-c}$ | Maximum charging capacity of storage $\xi$ |
| $\underline{p}_{\xi}^{ess-c}$ | Minimum charging capacity of storage $\xi$ |
| $\overline{q}_{i,g,t}^{imp}$ | Upper limit of imports block *g* for border *i* |
| $\overline{q}^{exp}_{i,g,t}$ | Upper limit of exports block $g$ for border $i$ |
| $\overline{q}_{u,t,m}$ | Upper limit of block $m$ of the linearized cost curve for unit $u$ |
| $rc^{e\pm}_{u}$ | Reserves capability for type $e$ and unit $u$ |
| $rc^{e\pm}_{\xi}$ | Reserves capability for type $e$ and storage $\xi$ |
| $rr^{e\pm}_{t}$ | System reserves requirements of type $e$ at $t$ |
| $T_{u,\gamma}$ | Time unit $u$ remains offline before going into longer standby conditions $\gamma$ |
| $T^{syn}_{u,\gamma}$ | Duration synchronization phase for unit $u$ for startup type $\gamma$ |
| $T^{soak}_{u,\gamma}$ | Duration soak phase for unit $u$ for startup type $\gamma$ |
| $T^{des}_{u}$ | Duration desynchronization phase for unit $u$ for startup type $\gamma$ |
| $T^{stop}_{u}$ | Duration of minimum down time for unit $u$ |
| $T^{run}_{u}$ | Duration of minimum up time for unit $u$ |
| $\overline{\varepsilon}_{\xi}$ | Maximum storage levels of $\xi$ |
| $\underline{\varepsilon}_{\xi}$ | Minimum storage levels of $\xi$ |
| $\overline{\varepsilon}^{ohps}_{u,t}$ | Maximum storage levels of open-loop HPS $u$ |
| $\underline{\varepsilon}^{ohps}_{u,t}$ | Minimum storage levels of open-loop HPS $u$ |
| $\upsilon_{u}$ | Ramp up rate of unit $u$ |
| $\varphi^{ens}$ | Penalty factor for the violation of active power balance equilibrium |
| $\varphi^{e\pm}$ | Penalty factor for the violation of reserves requirements equilibrium $e$ |